\documentclass[a4paper,fleqn,usenatbib]{mnras}

\usepackage[T1]{fontenc}
\usepackage{ae,aecompl}

\usepackage{graphicx}	
\usepackage{amsmath}	
\usepackage{amssymb}	
\usepackage{lipsum}
\usepackage{subfigmat}
\graphicspath{{./Figures/}}
\usepackage{enumerate}
\usepackage{titlesec}
\titlespacing*{\section}{0.5pt}{1.1\baselineskip}{\baselineskip}
\titlespacing*{\subsection}{0.5pt}{1.1\baselineskip}{\baselineskip}
\titlespacing*{\subsubsection}{0.5pt}{1.1\baselineskip}{\baselineskip}

\title[The structure of young embedded protostellar discs]{The structure of young embedded protostellar discs}

\author[B. A. MacFarlane et al.]{
Benjamin A. MacFarlane,$^{1}$\thanks{E-mail: bmacfarlane@uclan.ac.uk}
Dimitris Stamatellos,$^{1}$
\\
$^{1}$Jeremiah Horrocks Institute for Mathematics, Physics and Astronomy, University of Central Lancashire, Preston, PR1 2HE, UK\\
}

\date{Accepted 2017 July 31. Received 2017 July 31; in original form 2017 April 08}

\pubyear{2017}

\begin{document}
\label{firstpage}
\pagerange{\pageref{firstpage}--\pageref{lastpage}}
\maketitle

\begin{abstract}
Young protostellar discs provide the initial conditions for planet formation. The properties of these discs may be different from those of late-phase (T Tauri) discs due to continuing infall from the envelope and protostellar variability resulting from irregular gas accretion. We use a set of hydrodynamic simulations to determine the structure of discs forming in collapsing molecular clouds. We examine how radiative feedback from the host protostar affects the disc properties by examining three  regimes: without radiative feedback, with continuous radiative feedback and with episodic feedback, similar to  FU Ori-type outbursts. We find that the radial surface density and temperature profiles vary significantly as the disc accretes gas from the infalling envelope. These profiles are sensitive to  the presence of spiral structure, induced by gravitational instabilities, and the radiative feedback provided by  the protostar, especially in the case when the feedback is episodic. We also investigate whether mass estimates from position-velocity (PV) diagrams are accurate for early-phase discs. We find that the protostellar system mass (i.e. the mass of the protostar and its disc) is underestimated by up to 20\%, due to the impact of an enhanced radial pressure gradient on the gas. The mass of early-phase discs is a significant fraction of the mass of the protostar, so position-velocity diagrams cannot accurately provide the mass of the protostar alone. The enhanced radial pressure gradient expected in young discs may lead to an increased rate of dust depletion due to gas drag, and therefore to a reduced dust-to-gas ratio. 
\end{abstract}




\section{Introduction}

Young embedded protostars (Class 0/I Young Stellar Objects, hereafter YSOs; \citealp{lada87, andre93}) offer a unique opportunity to study protostellar discs at their earliest stages of evolution. 

Much of our knowledge on properties of protostellar discs comes from theoretical and observational studies of T Tauri discs, i.e. non-embedded Class II/III objects. Theoretical and numerical works  provide predictions for the disc structure. Considering a passive disc in which only heating due to protostellar luminosity is present, for a geometrically thin disc it can be shown that $T (r) \propto r^{-3/4}$ \citep{adamshu86, kenhart87, chiangold97}. For passive discs where the vertical hydrostatic equilibrium is considered, the disc is expected to flare (i.e. the disc height $h$ increases with radius $r$). As a flared disc absorbs a larger fraction of the protostellar radiation, \citet{kenhart87} find that $T(r) \propto r^{-1/2}$. Observations of  T Tauri discs show temperature profiles similar to these predictions \citep{beckwith90,osterlohbeckwith95,andrews09}, with  $T(r)  \propto r^{-q}$, where $q$ is between $0.35$ and $0.8$. Regarding the radial density structure, \citet{linpringle90}, through consideration of cloud collapse and subsequent disc formation, find that $\Sigma \propto r^{-p}$ where $p$ lies between $1$ and $3/2$. Observations by  \citet{andrews09} have shown that for the discs in the $\sim 1 \ \text{Myr}$-old Ophiucus star forming region the density drops as  $\Sigma \propto r^{-p}$, where  $p \sim0.4-1$.

The properties of  young discs embedded in their cloud cores (Class 0 objects) are more difficult to determine, due to the envelope dominating over disc emission at long wavelengths \citep{looney03, chiang08}. However, recent  observations have begun to detect discs of  Class 0/I YSOs \citep{tobin15}, with authors reporting signatures of Keplerian rotation in the gas for sufficiently well resolved objects (e.g. \citealp{tobin12, yen15}). Recently, ALMA observations have revealed early phase discs around HL Tau \citep{alma15}, Elias 2-27 \citep{perez16} and L1448 IRS3B \citep{Tobin:2016b}. Particularly for Elias 2-27 and L1448 IRS3B, such observations show that indeed discs can form at a very early stage and that they can be relatively massive  (a few tenths of the solar mass) and extended  (up to a few hundred AU). In support of the existence of massive early phase discs, \citet{dunham14} find that uncertainties associated with the assumed dust opacity and dust temperature can lead to disc masses being underestimated by a factor of 2-3 when using millimetre observations, and up to an order of magnitude when using sub-millimetre observations. Similar findings have been reported by \citet{evans17} who note that in self-gravitating discs the opacity and temperature may vary considerably throughout the disc, resulting in underestimated disc mass. \citet{birnstiel09, birnstiel10} also question the reliability of flux-based disc mass estimates. The authors argue that in a disc with a gas radial pressure gradient, radial migration and subsequent protostellar accretion of dust leads to a dust-to-gas ratio lower than is commonly assumed (typically $0.01$), resulting in underestimates of the total mass. 

The study of embedded protostellar discs are of particular interest, due to their propensity to become gravitationally unstable and fragment  to form massive planets, brown dwarfs and low-mass stars \citep{vorobyovbasu10a,machida11,vorobyov13,stamatelloswhitworth09,stamatellos:2011b,machida08,kratter10}. \citet{dunham14} argue that when mass estimates are corrected for, between 50-70\% of early phase discs may be in the regime where disc self-gravity is significant. It is therefore important to determine the properties of these discs (e.g. density and temperature profile) both observationally and theoretically.


Radiation-hydrodynamic (RHD) simulations (with or without magnetic fields) provide a way to study the properties of young discs as they form in collapsing cloud cores \citep{Commercon:2011d, Tomida:2013a, Bate:2014a, Tsukamoto:2015a, Tsukamoto:2015b}. \citet{Stamatellos:2011a} and \citet{stamatellos12} find that discs grow quickly in mass by accreting gas from the infalling envelope to become gravitationally unstable and fragment. The tendency for discs to become gravitationally unstable is characterised by the Toomre Q parameter \citep{toomre64}, defined as:
\begin{equation}\label{eq:toomreq}
	Q = \frac{c_{s} \Omega}{\pi G \Sigma},
\end{equation}
where $c_{s}$ is the local sound speed, $\Omega$ is the Keplerian frequency and $\Sigma$ is the local surface density. Previous work has found that the value of $Q$ required to form gravitational instabilities (GIs) and resultant spiral features is on the order of $Q \lesssim 1.4$ \citep[e.g.][]{laughlinboden94}. \citet{gammie01} found that for fragmentation to happen the disc needs to cool fast enough, i.e. $t_\text{cool} < (1-2) \ t_\text{orb}$, where $t_\text{orb}$ is the local orbital period.


The goal of this work is to determine the structure of discs in the Class 0/I phase using RHD simulations. More specifically, we examine how radiative feedback from the host protostar affects the properties of  young embedded discs. We also investigate how these properties may be conveyed from the kinematics of gas, as this is described in position-velocity diagrams.

The paper is structured as follows. In \S~\ref{ICs_Methods} we outline the radiation hydrodynamic simulations  we perform. We present the results regarding  the disc structure (radial surface density and temperature profiles) in  \S~\ref{pq_comp}. We provide analyses of mass estimates of the protostellar system through synthetic position-velocity diagrams in \S~\ref{pv_analysis}. In \S~\ref{conclusions}  we present the conclusions  of this work.

\begin{figure*}
 \includegraphics[width=150mm]{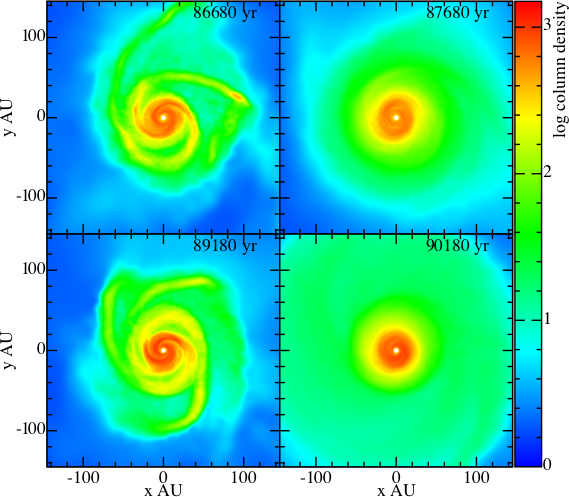}
 \caption{Column density ($\text{g cm}^{-2}$) plots (at different times as marked on each graph) of typical disc morphologies for the non radiative feedback run (NRF - top left), the continuous radiative feedback  run (CRF - top right) and the episodic radiative feedback  run A (pre-outburst ERF-A - bottom left ; during-outburst ERF-A - bottom right). }
 \label{fig:spatial}
\end{figure*}

\section{Initial conditions and computational methods}\label{ICs_Methods}

We make use of simulations presented in \citet{stamatellos12}. These simulations, performed using the SPH code SEREN \citep{hubber11a, hubber11b}, follow the collapse of a $5.4 \ \text{M}_\odot$ pre-stellar core over $100 \ \text{kyr}$. The initial density profile of the core is set to
\begin{align}
\rho (r) = \frac{\rho_{\text\tiny{C}}}{\left[1+(r/R_{\text\tiny{C}})^{2}\right]^{2}},
\end{align}
where $\rho_{\text\tiny{C}} = 3\times 10^{-18} \ \text{g cm}^{-3}$ is the central density, and $R_{\text\tiny{C}} = 5000 \ \text{AU}$ is the radius in which the radial density profile flattens out. The core radius is $R = 50 \ 000 \ \text{AU}$. The pre-stellar core is assumed to be isothermal at the beginning of the simulation, with $T = 10 \ \text{K}$. These are typical values for observed prestellar cores \citep[e.g.][]{Andre:2014a}. Each simulation uses $10^{6}$ SPH particles, with each particle having mass $m_\text{SPH}= 5.4\times10^{-6} \ \text{M}_\odot$. The minimum resolvable mass is $M_\text{min} \sim m_\text{SPH} N_\text{neigh} \sim 3\times10^{-4} \ \text{M}_\odot$, where the number of SPH neighbours is assumed to be $N_\text{neigh} = 50$. This mass ensures that all self-gravitating condensations are well resolved \citep{whitworthstamatellos06}. The code makes use of a time-dependent artificial viscosity \citep{morrismonaghan97}  with parameters $\alpha_{\min}=0.1$, $\alpha_{\max}=1$ and $\beta=2\alpha$ to reduce the artificial shear velocity. Artificial viscosity is adopted to parameterise angular momentum transport due to disc turbulence and/or the mangeto-rotational instability. This ensures that disc viscous evolution occurs even when the disc is not susceptible to angular momentum transport through GIs.

The simulations employ a selection of radiative feedback treatments from the protostar to study their effect on disc evolution. SPH particle density, temperature and gravitational potential is used to estimate particle mean optical depth, which is used to solve the energy equation for each SPH particle at each time step. A pseudo-background radiation field is imposed throughout the disc, so that
\begin{align}
T(r) = \left[(10 \ \text{K})^{4} + \frac{L_*}{16 \pi \sigma_\text{SB} r^{2}}\right]^{1/4}.
\end{align}
This temperature acts as a floor below which material cannot radiatively cool. $\sigma_\text{SB}$ is the Stefan-Boltzmann constant, $L_*$ is the luminosity of the central protostar and $r$ is the distance from the protostar. For a protostar with radius $R_*$ and mass $M_*$, the luminosity is:
 \begin{align}
 L_* = \bigg ( \frac{M_*}{M_\odot}\bigg ) ^{3}L_\odot + f \frac{G M_* \dot{M}}{R_*}\,,
 \end{align}
where $\dot{M}$ is the accretion rate onto the protostar and $f~=~0.75$ is the efficiency of gravitational potential energy release during accretion. Disc material is heated primarily due to  the protostellar accretion luminosity, as well  due to the intrinsic protostellar luminosity and hydrodynamic processes (i.e. compressive and viscous heating). Atomic and molecular processes, including dissociation of H$_2$, ionisation of atomic H, rotational and vibrational degrees of freedom of H$_2$ and opacity changes due to melting of ice on dust grains and sublimation of dust  are taken into account. The authors direct the interested reader to \citet{stamatellos07} for further details on the radiative transfer method. A key feature of the numerical method of \citet{stamatellos12} is the inclusion of different regimes of radiative feedback from the central protostar.  We discuss these in detail in the following paragraphs. 

\section{Hydrodynamic simulations of cloud collapse and disc formation}

Five runs were performed that employ different radiative feedback from the central protostar (see Table~\ref{tab:ERFparams}). The simulations are therefore identical until the formation of the first protostar, and diverge afterwards. The protostar forms after $\sim 79 \ \text{kyr}$, and a disc forms thereafter and   grows due to mass accretion from the surrounding envelope. For the entirety of the simulations, the mass of the envelope is significantly greater than the protostar and disc, i.e. our analyses focus on the study of deeply embedded  YSOs. 

\subsection{Simulation I: No radiative feedback (NRF)}
	
In this case, although material accretes onto the protostar, heating of disc material takes place only through compression and viscous dissipation. This model (which we henceforth refer to as NRF) represents a benchmark from which the effects of protostellar luminosity on disc structure can be evaluated.

\citet{stamatellos12} report that due to the lack of radiative heating there is a  sharp decrease in disc temperature as a function of radius. Additionally, due to mass loading from the envelope onto the disc, strong GIs develop with  a few kyr after the formation of the protostar. GIs subsequently result in disc fragmentation and formation of stellar mass  companions. Continued fragmentation and companion formation leads to multiple stellar mass objects, significantly disrupting the protostellar disc. We present a snapshot of the NRF model prior to disc fragmentation in Fig.~\ref{fig:spatial} (top left panel) to indicate the prevalence of GI features.

	\subsection{Simulation II: Continuous radiative feedback (CRF) }

The second case we examine uses continuous feedback (CRF) from the protostar. Therefore, energy released due to protostellar accretion and hydrodynamic processes heats the protostellar disc. As matter flows continuously from the disc to the protostar, the luminosity of the central protostar remains high ($\sim 20-100 \ \text{L}_\odot$). 

\citet{stamatellos12} find that, in agreement with \citet{bate09} and \citet{offner09}, continuous radiative feedback acts to suppress disc fragmentation. Although a disc-to-star mass ratio of $\sim 1$ for a long period during the evolution of the system would indicate that strong GIs should persist in the disc, GIs are generally suppressed due to the high disc temperature resulting in $Q\gtrsim1$ at all radii. \citet{stamatellos12} find that the disc does become marginally unstable at some points during the evolution of the disc, however the resultant increase of protostellar accretion due to the GIs increases radiative heating, leading to self-regulation of the GI spiral structures \citep{lodatorice04, lodatorice05}. In this run the disc does not fragment.

	\subsection{Simulations III-V: Episodic radiative feedback (ERF)}

In the third  case we examine, the protostar accretes matter episodically resulting in  episodic radiative feedback (ERF).
Episodic radiative feedback has been observed in late-phase discs \citep[FU Ori and EXor type outbursts, e.g.][]{Liu:2016b} and has been proposed as a possible solution to the luminosity problem, i.e. the fact that the observed  luminosities of young protostars are much lower than expected from simple theoretical arguments relating to the final stellar mass and the time needed to accumulate this mass \citep[see][and references therein]{Audard:2014a}. Such outbursts are believed to happen more frequently and be more prominent in younger protostars \citep[e.g.][]{Vorobyov:2015a} but only one Class 0 outbursting object has been been observed so far \cite{Safron:2015a}. \cite{Johnstone:2013a} examined the effect of an outburst from an embedded protostar on the thermal properties of its envelope  and found that the overall spectral energy distribution of the object responds to the change of the protostellar luminosity within a relatively short time after the outburst (up to a few months).

 In the model presented here, episodic accretion occurs through angular momentum transport due to GIs in the outer disc, and the magneto-rotational instability (MRI) in the inner disc \citep{armitage01, zhu09a, zhu09b, zhu10a, zhu10b, Stamatellos:2011a, Mercer:2017a}.  During the quiescent phase (i.e. when there is no accretion event), radiative feedback is primarily due to intrinsic protostellar luminosity. In this phase, the luminosity of the protostar is $\lesssim 1 \ \text{L}_\odot$. This relatively low protostellar luminosity  is not sufficient to prevent GIs in the outer disc. These GIs redistribute angular momentum outward, allowing matter to flow radially inward. Matter is not accreted onto the protostar, as GIs are not present in the inner disc region due to the high temperature. Instead, matter accumulates in the inner disc region ($1 \ \text{AU}$) around the central protostar. As the mass reservoir of the inner disc region increases, radiative, compressive and viscous heating raise the temperature. When the temperature exceeds the MRI activation threshold of $T_\text{\tiny{MRI}} \sim 1400 \ \text{K}$ \citep{zhu09a, zhu09b, zhu10a, zhu10b}, an effective MRI viscosity, $\alpha_\text{\tiny{MRI}}$, redistributes angular momentum in the inner disc. The net effect is that all matter from the inner disc is accreted to the central protostar. We characterise this period as the outburst phase. The accretion luminosity during the outburst phase is up to $\sim 5$ orders of magnitude greater than during the quiescent phase, for a typical outburst duration of over $\sim \text{few} \ 100 \ \text{yr}$. Once the outburst phase has started, disc is heated sufficiently and therefore it is stable. The disc remains heated during the outburst, but a few kyr after the outburst stops, it cools and continued mass accretion from the surrounding envelope leads to GIs. 

The ERF simulations presented here are parameterised by the accretion rate on to the protostar in the quiescent phase, $\dot{M}_{\text{\tiny{Q}}}$, and the viscous alpha parameter $\alpha_\text{\tiny{MRI}}$ \citep{ss73} for the MRI-active inner disc.  Models with higher $\alpha_\text{\tiny{MRI}}$ exhibit shorter, more intense outburst phases. An increase of $\alpha_\text{\tiny{MRI}}$ results in quicker angular momentum transfer and a shorter timescale for the inner disc to deplete its mass reservoir onto the protostar. In this work, we focus on three runs: ERF-A, ERF-B and ERF-C respectively.  Parameters for each model are noted in Table~\ref{tab:ERFparams}.

We note that in the ERF-B and ERF-C models, akin to the NRF model, the disc fragments to form companions to the central protostar, with masses  at the end of the simulation greater that the deuterium-burning mass limit ($> 13 \ M_\text{J}$; \citealt{spiegel11}). Fig.~\ref{fig:spatial} demonstrates the typical surface density distribution of the ERF models, in the quiescent (bottom left) and outbursting (bottom right) phases for the ERF-A model.

\section{Density and temperature structure of embedded protostellar discs}\label{pq_comp}

\begin{table*}
	\centering
	\caption{Characteristic parameters of the 5 runs that we analyse in this work. $\alpha_\text{\tiny{MRI}}$ denotes the viscosity for the MRI-active inner disc, $\dot{M}_{\text{\tiny{Q}}}$ is the rate of mass accretion onto the protostar during the quiescent phase. $t_\text{i}$ and $M_\text{f}$ denote the formation time and final mass at the end of the simulation ($100 \ \text{kyr}$) of the central star (first value) and the secondary objects formed by disc fragmentation.}
	\begin{tabular}{lcccccr} 
		\hline
	 Simulation & Model (abbr.) & $\alpha_\text{\tiny MRI}$ & $\dot{M}_{\text{\tiny{Q}}} \ (M_\odot \ \text{yr}^{-1})$ & $t_\text{i} \ (\text{kyr}) $ & $M_\text{f} \ (M_\odot)$\\
		\hline
		\hline
		No radiative feedback & NRF & - & - & $78.7, 87.2, 88.8, 90.0$ & $0.49, 0.32, 0.005, 0.23$ \\
		Continuous radiative feedback & CRF & - & - & 78.7 & 0.59 \\
		Episodic radiative feedback: Run A & ERF-A & $0.01$ & $10^{-7}$ & 78.7 & 0.60 \\
		 Episodic radiative feedback: Run B & ERF-B & $0.3$ & $10^{-7}$ & 78.7, 90.7, 91.0 & 0.63, 0.26, 0.12 \\
		 Episodic radiative feedback: Run C & ERF-C & $0.1$ & $10^{-8}$ & 78.7, 89.0, 90.0 & 0.56,0.35, 0.11 \\
		\hline
	\end{tabular}
	\label{tab:ERFparams}
\end{table*}

\begin{figure*}
 \includegraphics[width=150mm]{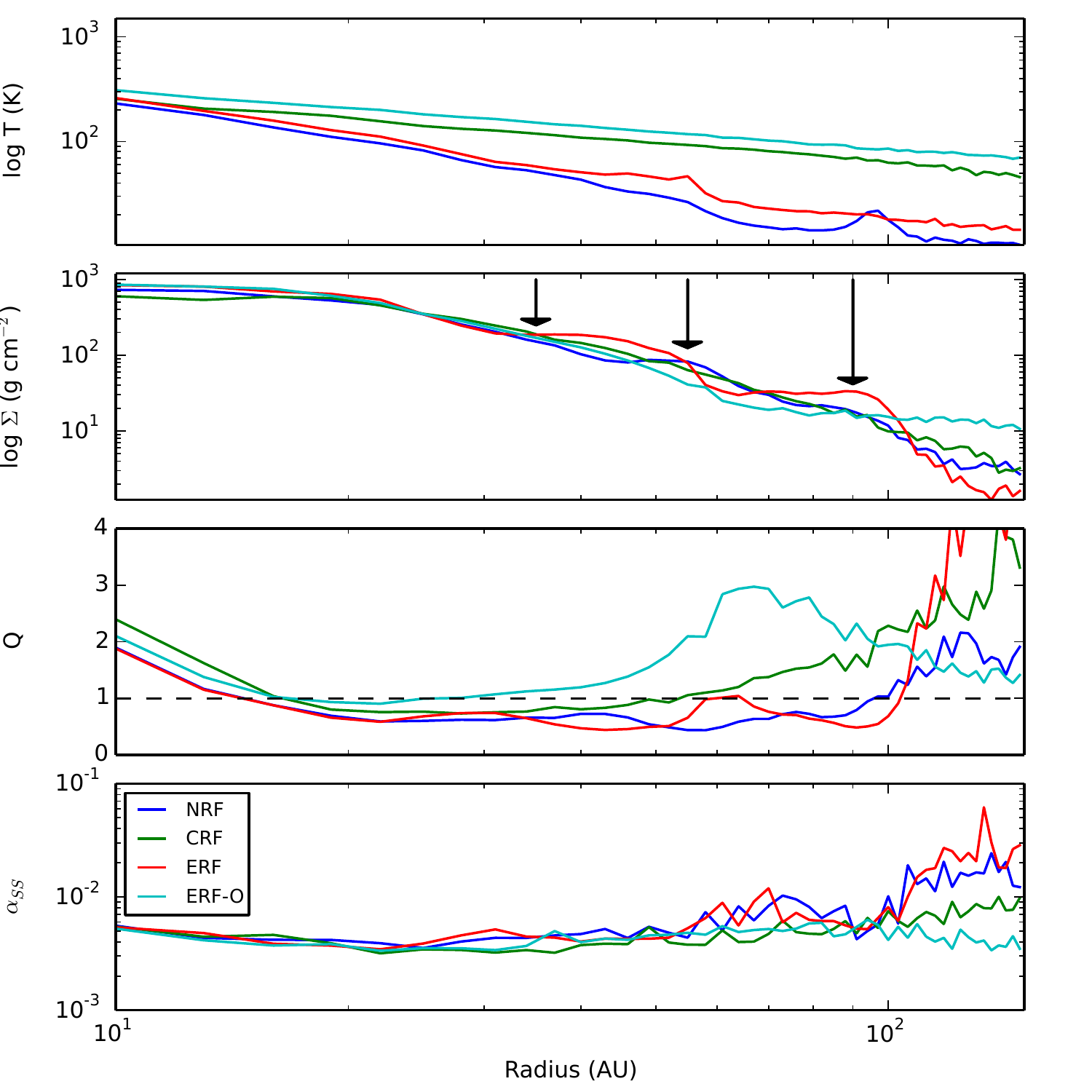}
 \caption{Azimuthally averaged radial profiles of the snapshots presented in Fig.~\ref{fig:spatial}. From top to bottom: radial profiles for temperature, surface density, Toomre Q parameter, and corresponding \citet{ss73} viscosity parameter. Each panel shows runs with no radiative feedback (NRF - blue line), with continuous radiative feedback (CRF - green line), and  with episodic radiative feedback run A (ERF: quiescent phase - red line ; ERF-O: outburst phase - cyan line).}
 \label{fig:AziProf_comp}
\end{figure*}
	
\begin{figure*}
 \centering
   \subfigure[NRF]{\includegraphics[width=58mm,keepaspectratio]{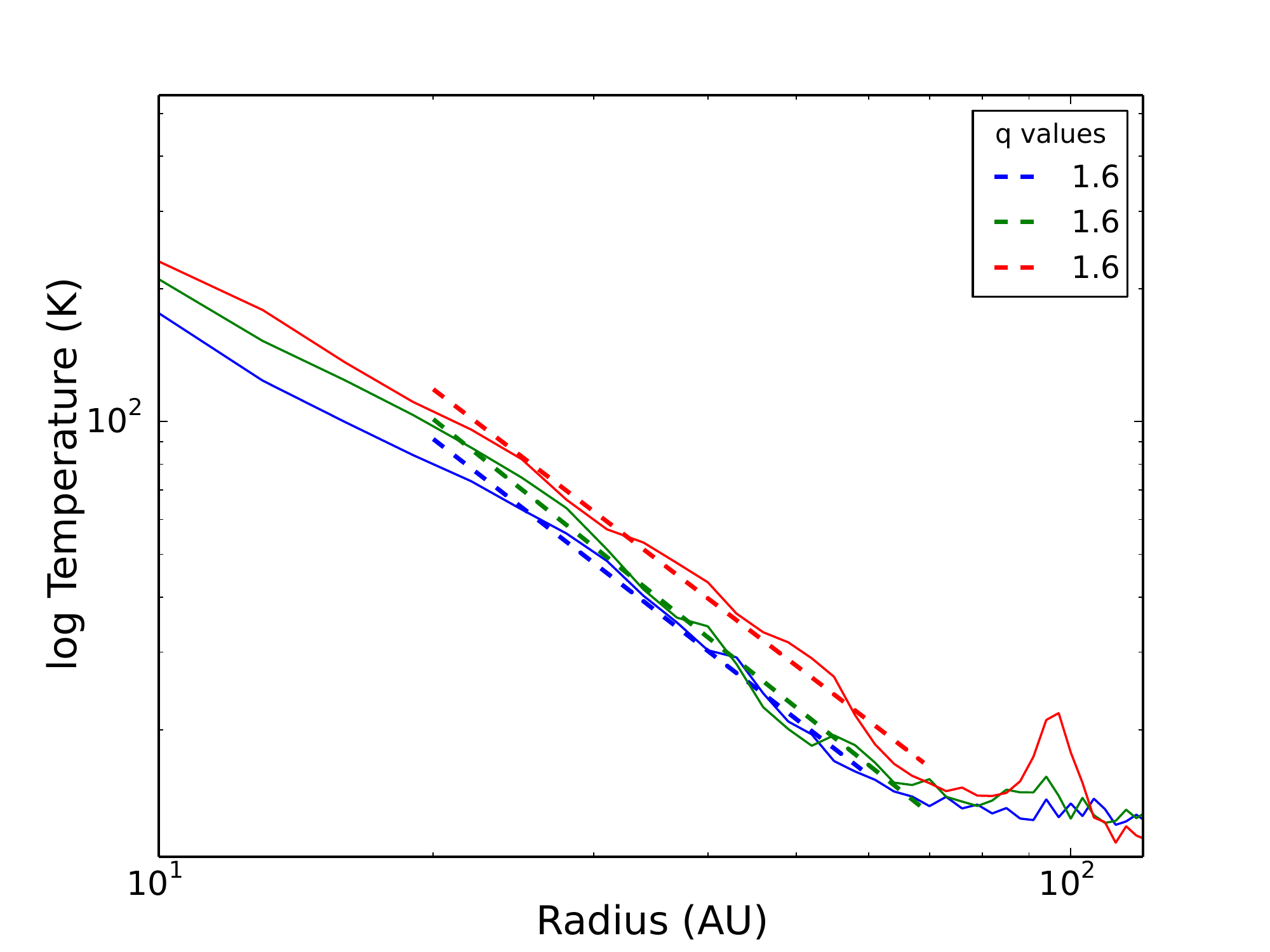}}
   \subfigure[CRF]{\includegraphics[width=58mm,keepaspectratio]{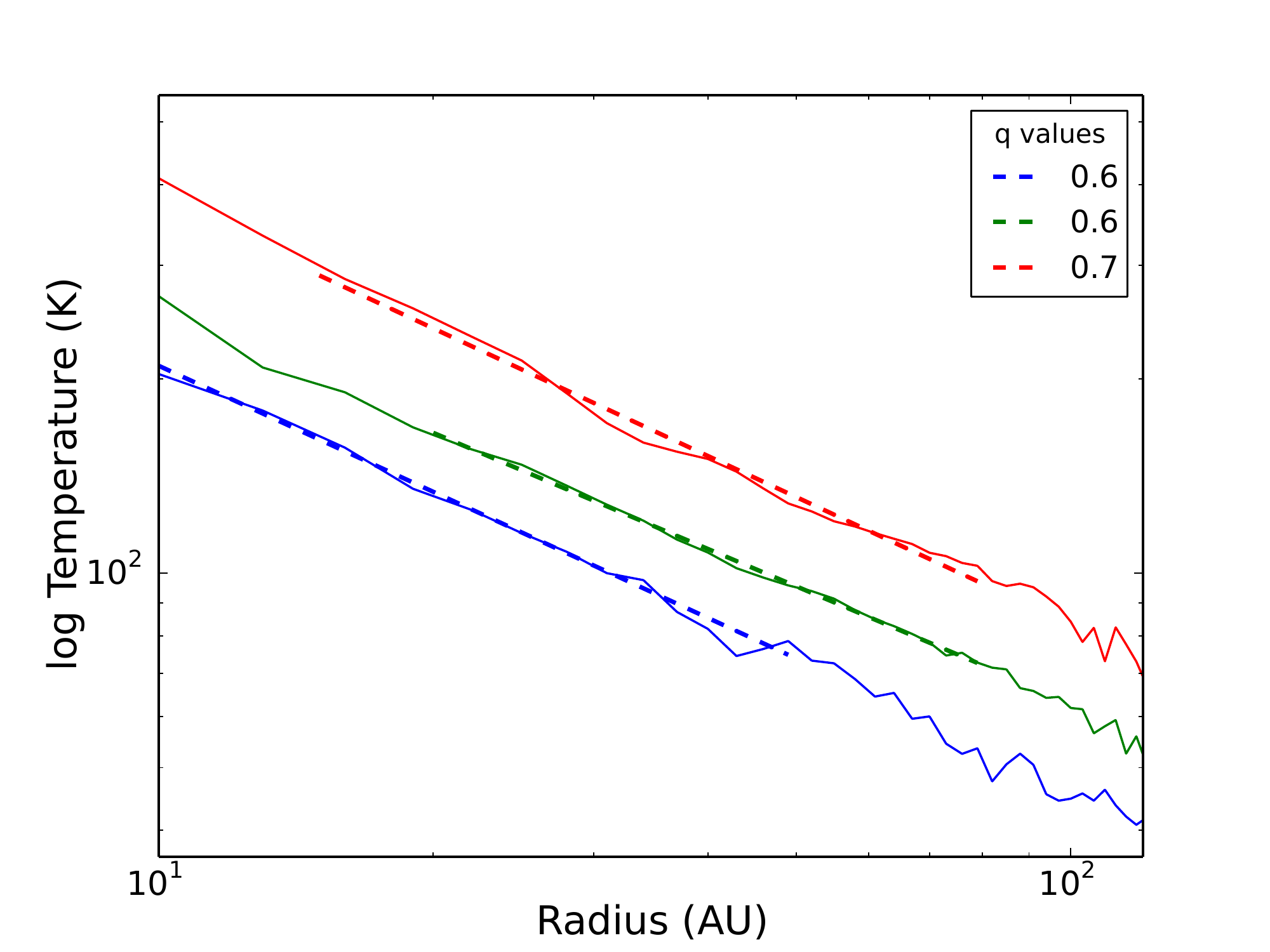}}
   \subfigure[ERF-A]{\includegraphics[width=58mm,keepaspectratio]{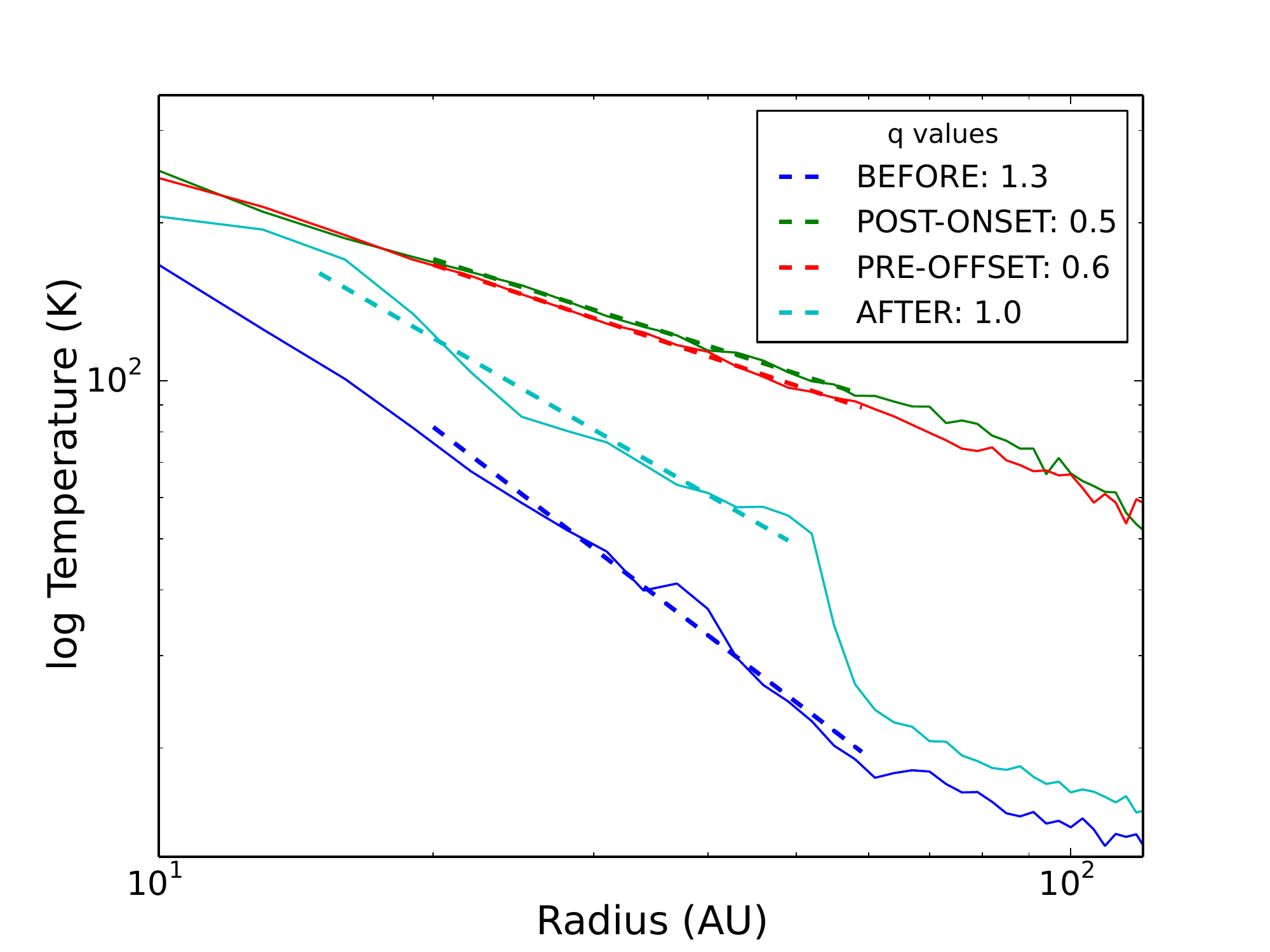}}
 \caption{Azimuthally averaged temperature profiles, with fits plotted as dashed lines. In the no radiative feedback (NRF) and continuous radiative feedback (NRF) case, 3 snapshots are plotted (blue, green, red corresponding to progressively later times). The estimated $q$ value is noted in each panel legend. For the episodic radiative feedback (ERF) run we plot 4 snapshots corresponding to the quiescent accretion phase (blue), the outburst phase just after it starts (post-onset; green), the outburst phase just before it terminates (pre-offset; blue), and finally after it terminates (post-offset; cyan).  ERF snapshot times and estimated $q$ values are also shown.}
 \label{fig:T_R}
\end{figure*}
	
\begin{figure*}
 \centering
   \subfigure[NRF]{\includegraphics[width=58mm,keepaspectratio]{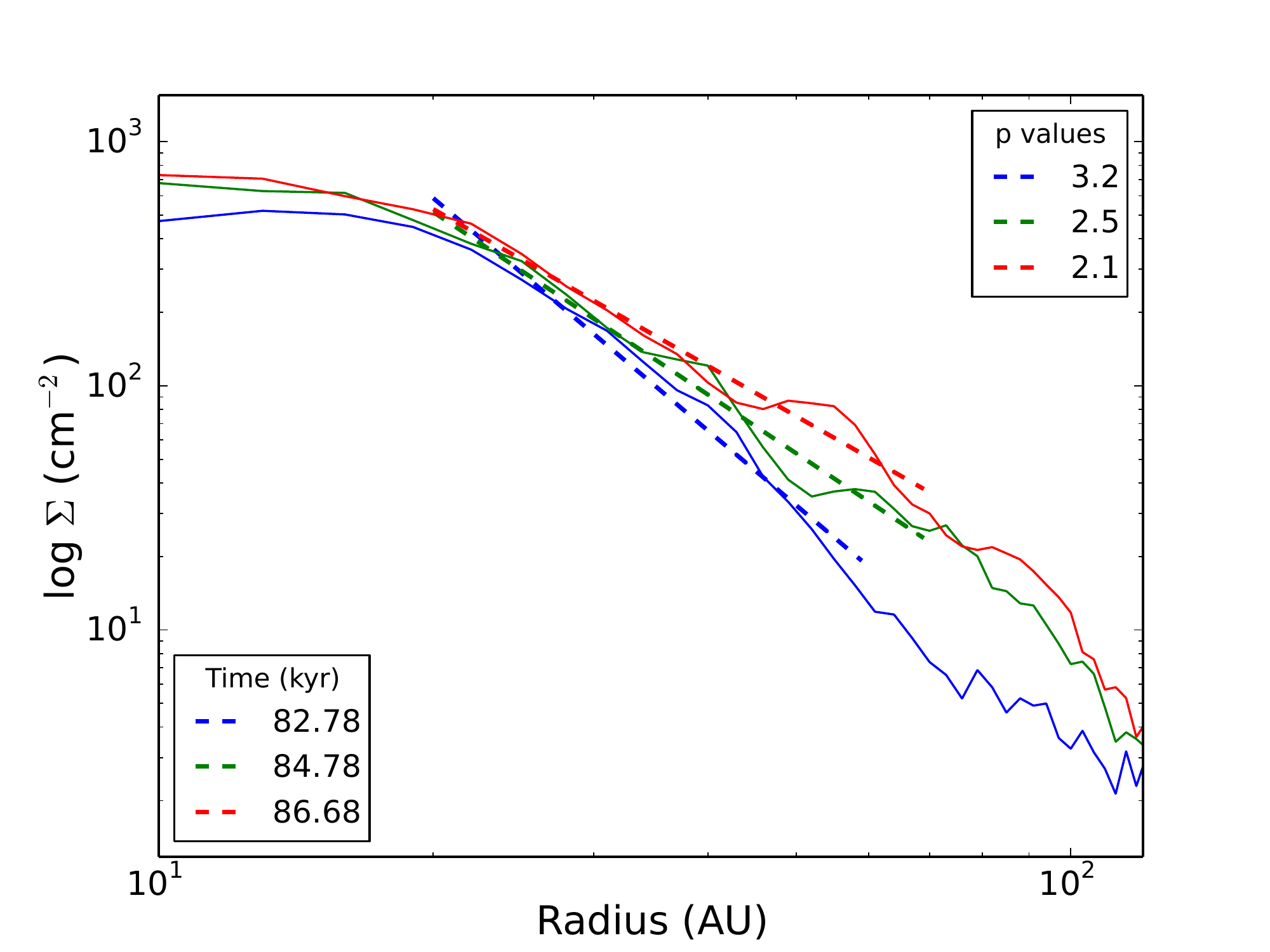}}
   \subfigure[CRF]{\includegraphics[width=58mm,keepaspectratio]{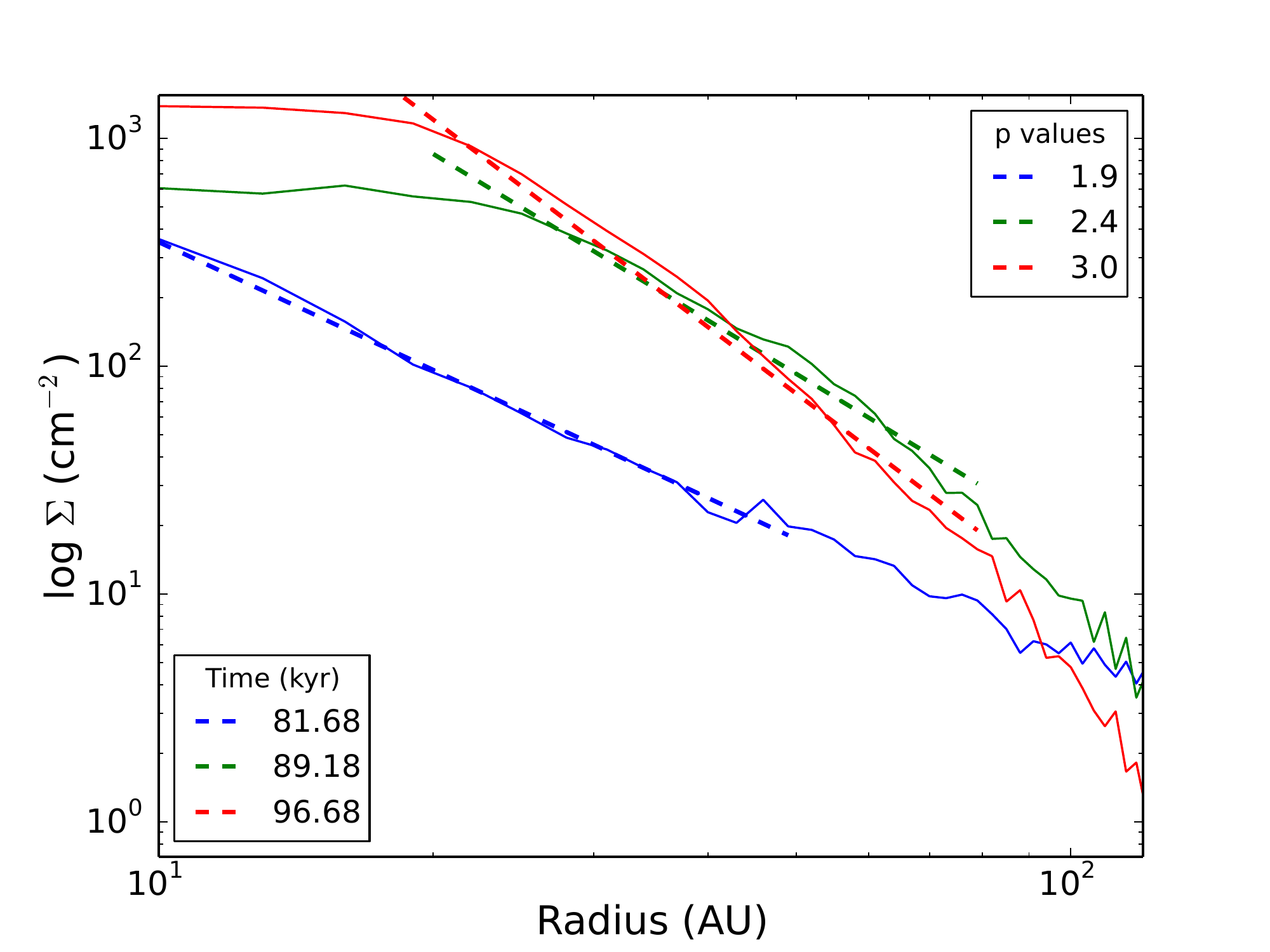}}
   \subfigure[ERF-A]{\includegraphics[width=58mm,keepaspectratio]{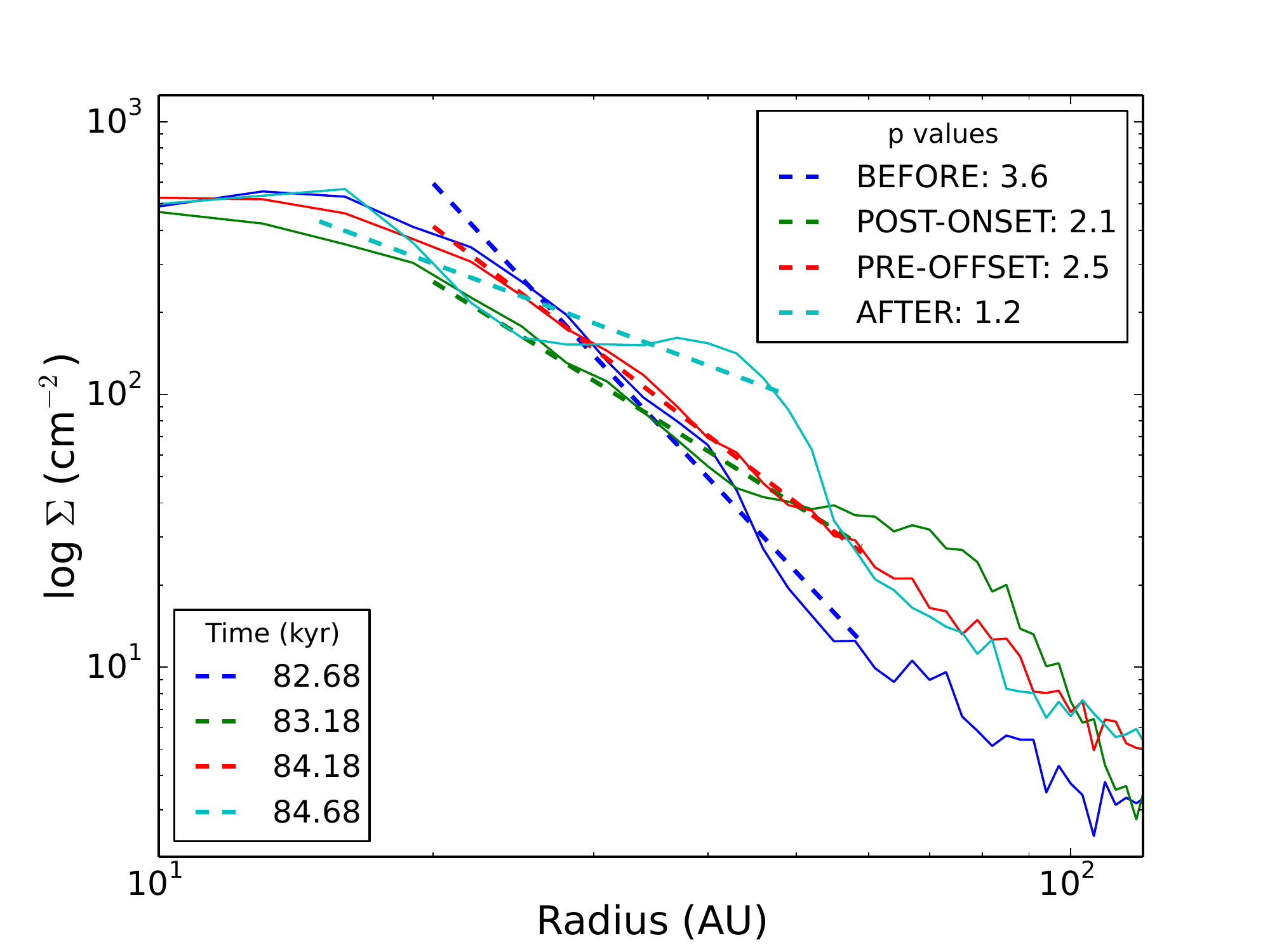}}
 \caption{As per Fig.~\ref{fig:T_R} but for azimuthally averaged surface density profiles. The estimated $p$ value is noted in the upper-right legend, with the time of the snapshot indicated in the bottom-left legend for each panel.}
 \label{fig:sig_R}
\end{figure*}
	
\begin{figure*}
 \centering
   \subfigure[NRF]{\includegraphics[width=58mm,keepaspectratio]{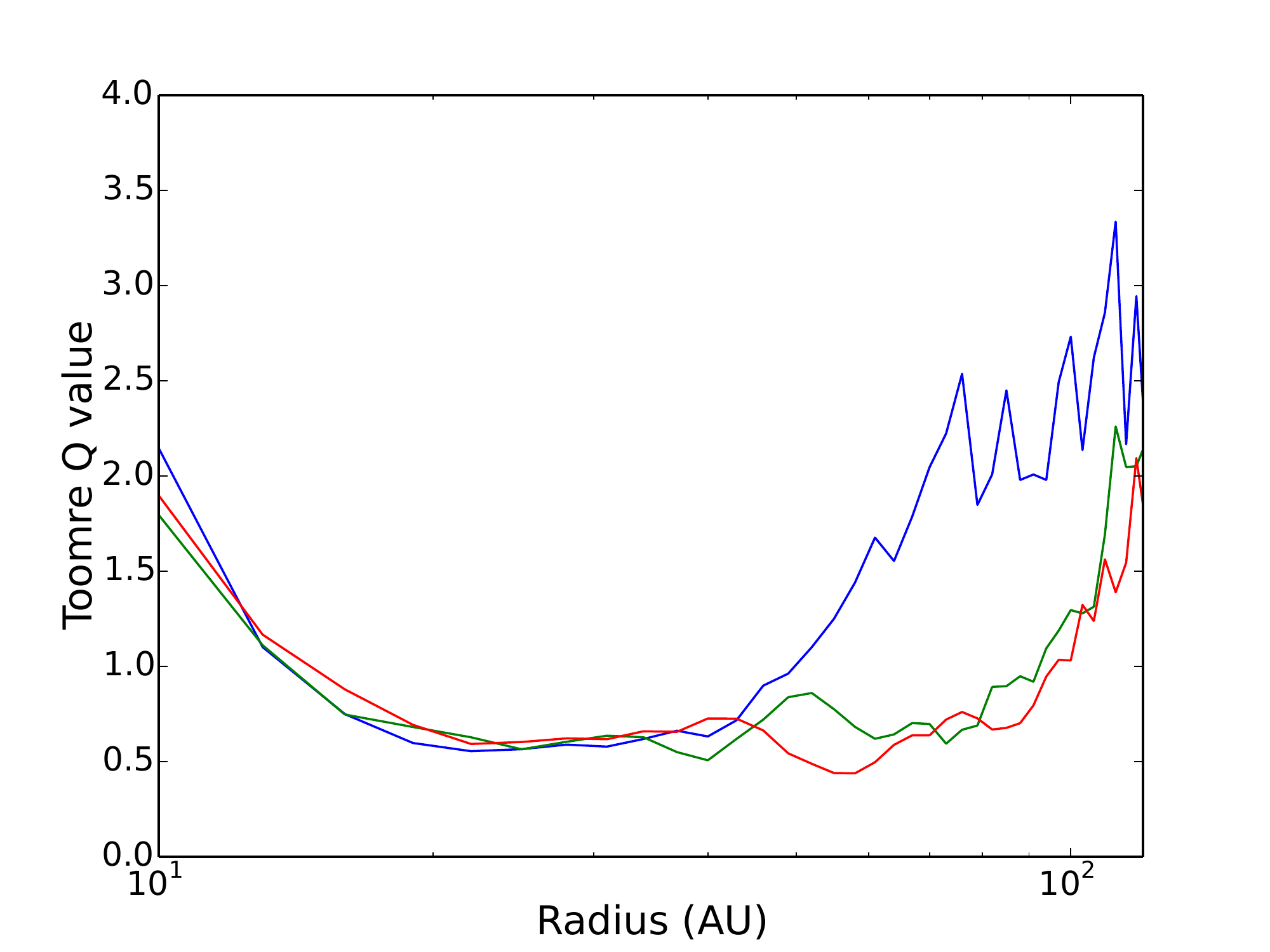}}
   \subfigure[CRF]{\includegraphics[width=58mm,keepaspectratio]{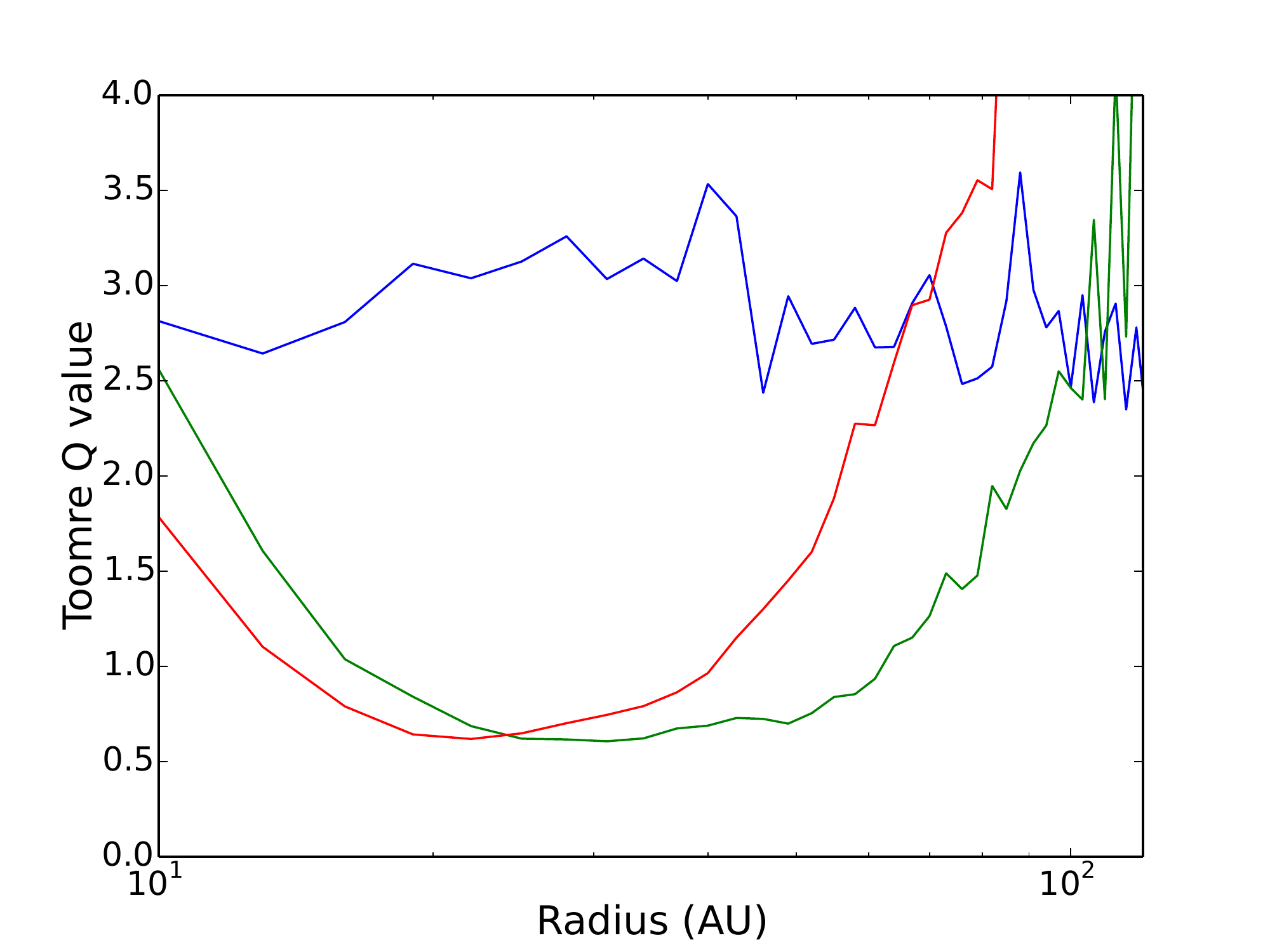}}
   \subfigure[ERF-A]{\includegraphics[width=58mm,keepaspectratio]{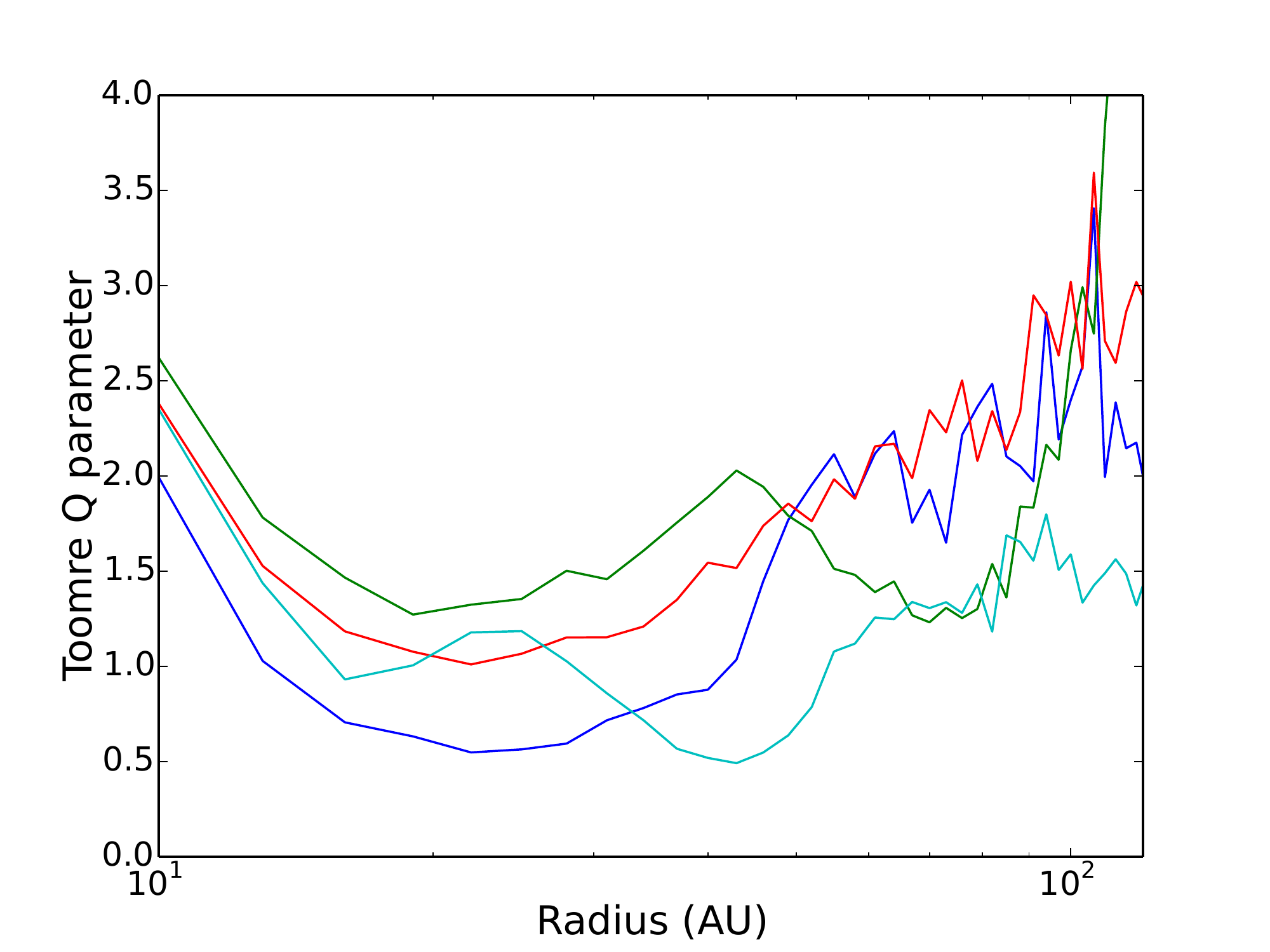}}
 \caption{As per Fig.~\ref{fig:T_R}, for Toomre Q radial distribution.}
 \label{fig:Q_R}
\end{figure*}

\begin{figure*}
 \centering
   \subfigure[NRF]{\includegraphics[width=58mm,keepaspectratio]{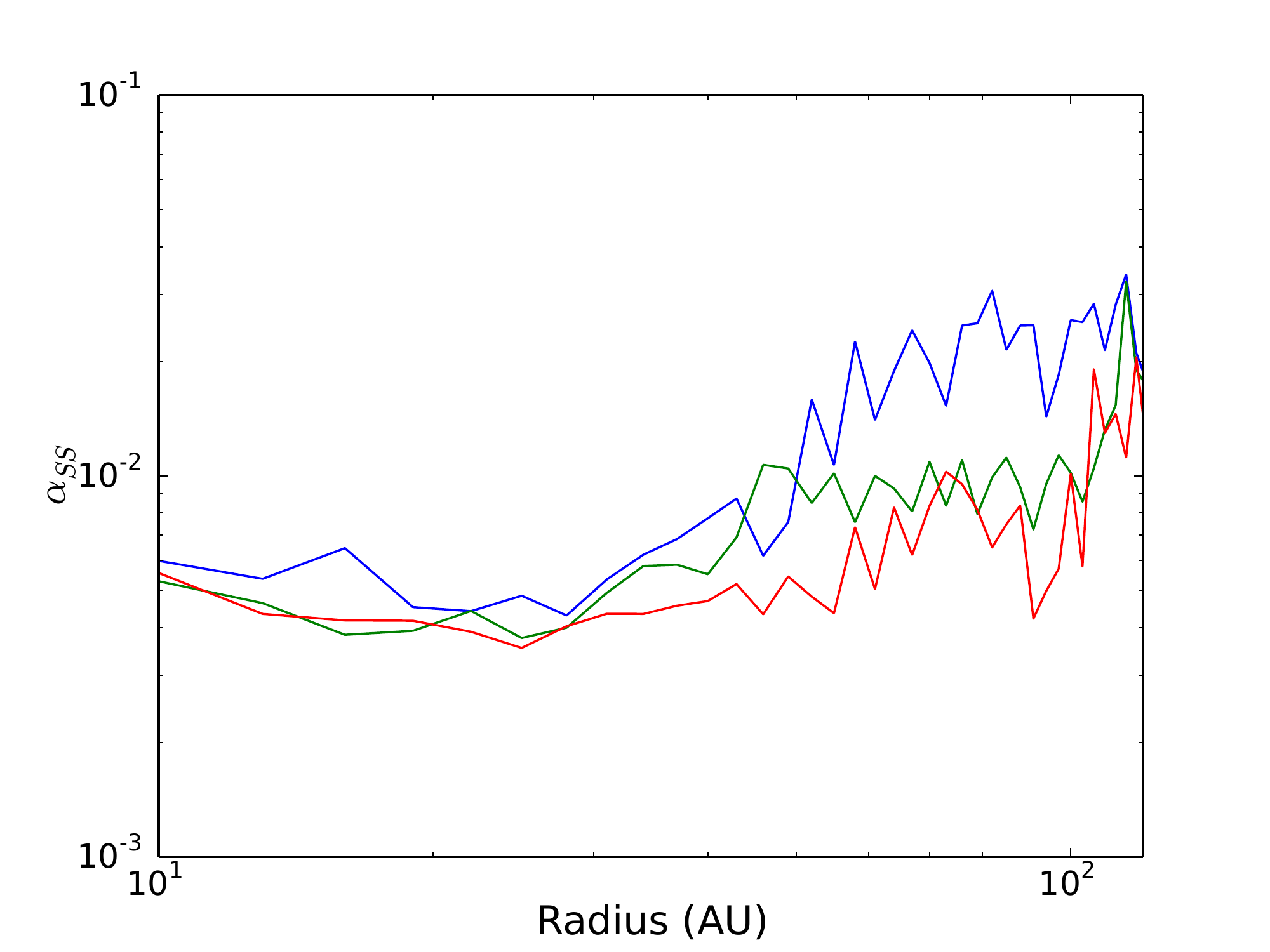}}
   \subfigure[CRF]{\includegraphics[width=58mm,keepaspectratio]{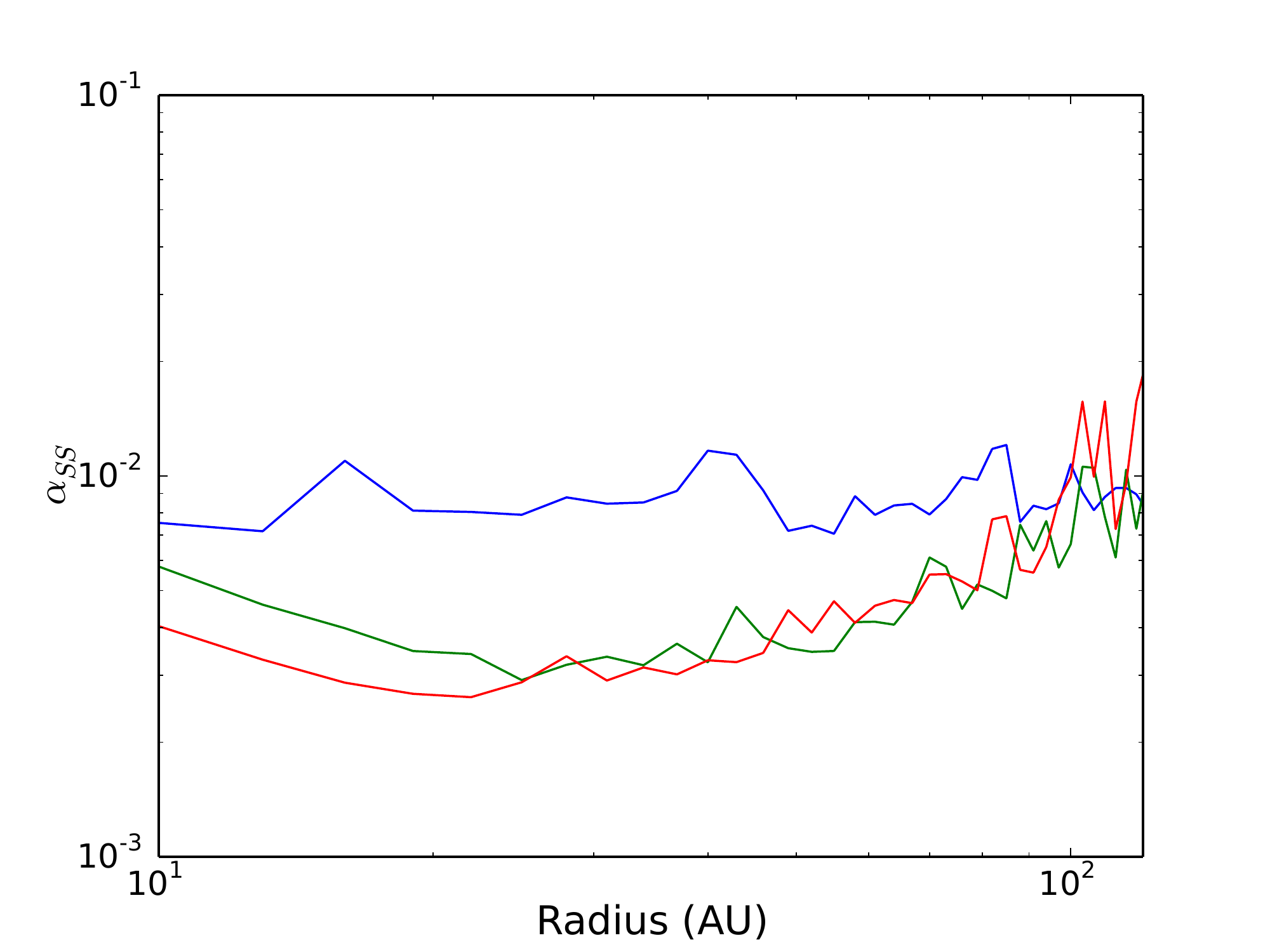}}
   \subfigure[ERF-A]{\includegraphics[width=58mm,keepaspectratio]{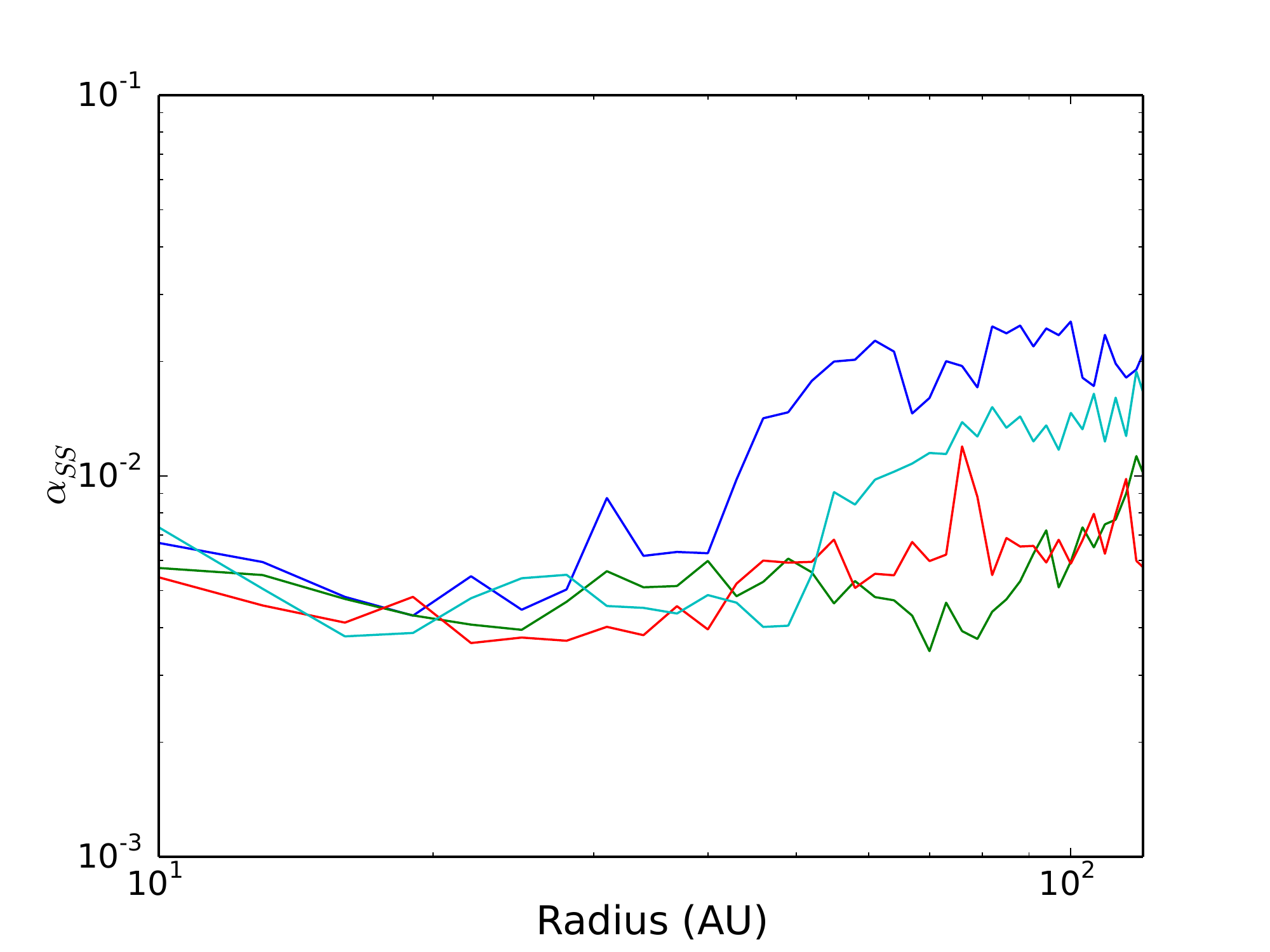}}
 \caption{As per Fig.~\ref{fig:T_R}, for the corresponding $\alpha_{SS}$ value of \citet{ss73}.}
 \label{fig:alpha_R}
\end{figure*}

We present azimuthally averaged temperature and surface density profiles in Fig.~\ref{fig:AziProf_comp}, for the snapshots of Fig.~\ref{fig:spatial}, comparing  the NRF, the CRF,  and  ERF-A  run (during both the quiescent and outbursting phase). Temperature profiles (top panel) demonstrate similarities between the NRF and the ERF-A during the quiescent accretion phase, whereas the  CRF model is quite similar with the ERF-A during the outburst accretion phase. These similarities can be explained due to the role of radiative feedback from the young protostar. For both the ERF-A/quiescent and NRF snapshots, radiative feedback is either weak or neglected, leading to lower temperatures when compared to both ERF-A/outburst and CRF snapshots. The disc in the  ERF-A run during the outburst  has higher temperature than in the CRF run. Surface density profiles demonstrate the  role of radiative feedback on shaping disc structure. In the NRF model, azimuthally averaged profiles retain signatures of GIs in the form of ``bumps'' at $r \gtrsim 50 \ \text{AU}$. Similar features are present for the quiescent phase ERF-A snapshot, however the bumps appear more distinct, suggesting stronger GIs. We also present radial Toomre Q profiles in Fig.~\ref{fig:AziProf_comp} (second from bottom panel) in order to evaluate whether the disc is gravitationally unstable and at what radii.  As expected, both the CRF and the outbursting ERF-A snapshots indicate that the disc is stable against GIs at all radii, whereas both NRF and quiescent ERF-A snapshots, with $Q \sim 0.5$ at large radial distances, suggest the presence of strong GIs. 
To evaluate the level of  SPH artificial viscosity, we adopt the method of \citet{lodatoprice10} \citep[see also][]{meru12}  to evaluate the azimuthally averaged viscosity parameter $\alpha_{SS}$ \citep{ss73}. This value is computed using
 \begin{align}
 \alpha_{SS} = \frac{1}{10} \alpha_{_{\rm SPH}} \frac{\langle h \rangle}{H},
 \end{align}
where $\langle h \rangle$ is the average smoothing length of SPH particles in the disc midplane (we assume that this is defined as $|z| < 0.05 R$). $H$ is the disc scale height.
%
%
We note that the coefficient in the above equation for $\alpha_{SS}$ (i.e. $1/10$) is sensitive to the choice of the SPH smoothing kernel (see \citealp{meru12} for details), however this does not significantly impact our findings. As a result of the numerical method, for the outer disc regions in which $Q > 2$, the angular momentum transport is due to SPH artificial viscosity, not GIs.

\begin{figure}
 \centering
 \includegraphics[width=70mm]{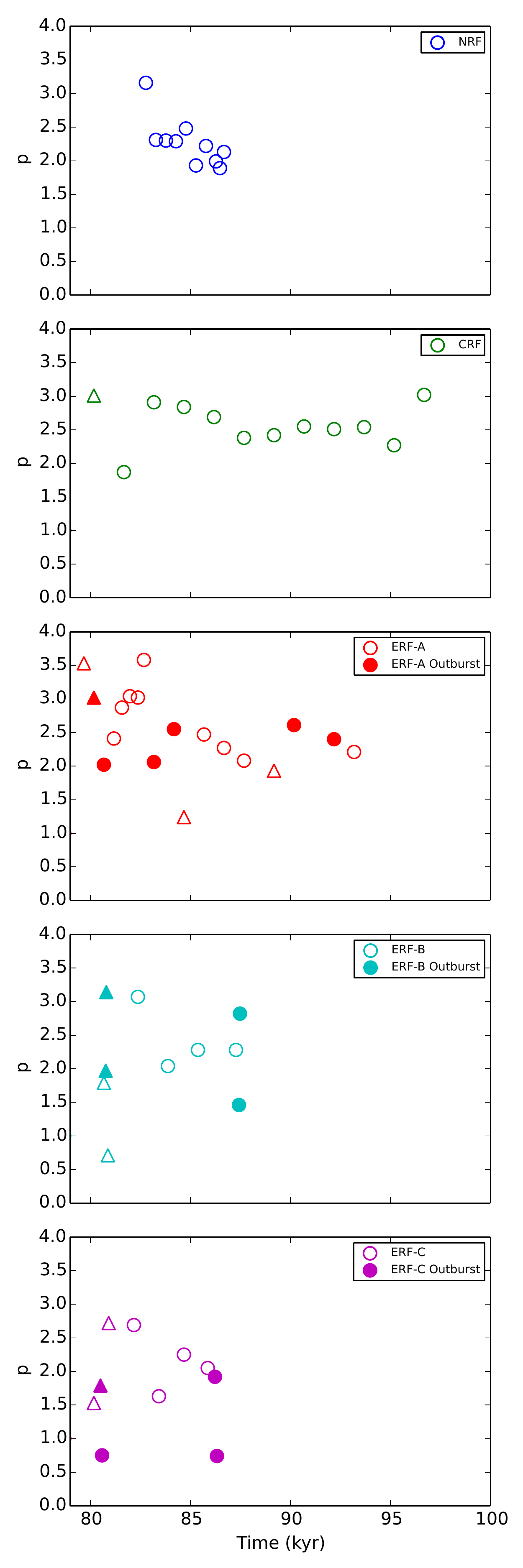}
 \caption{The evolution of the power exponent of  the surface density profile, $p$. No feedback (NRF), and run B and C of episodic feedback model (ERF-B and ERF-C) snapshots taken prior to disc fragmentation. Filled circles in ERF data represent times during accretion outbursts. Triangles indicate fits with increased uncertainty due to either (a) narrow radial range fit or (b) the existence of a central bulge in the disc.}
 \label{fig:p_Time}
\end{figure}

\begin{figure}
 \centering
 \includegraphics[width=70mm]{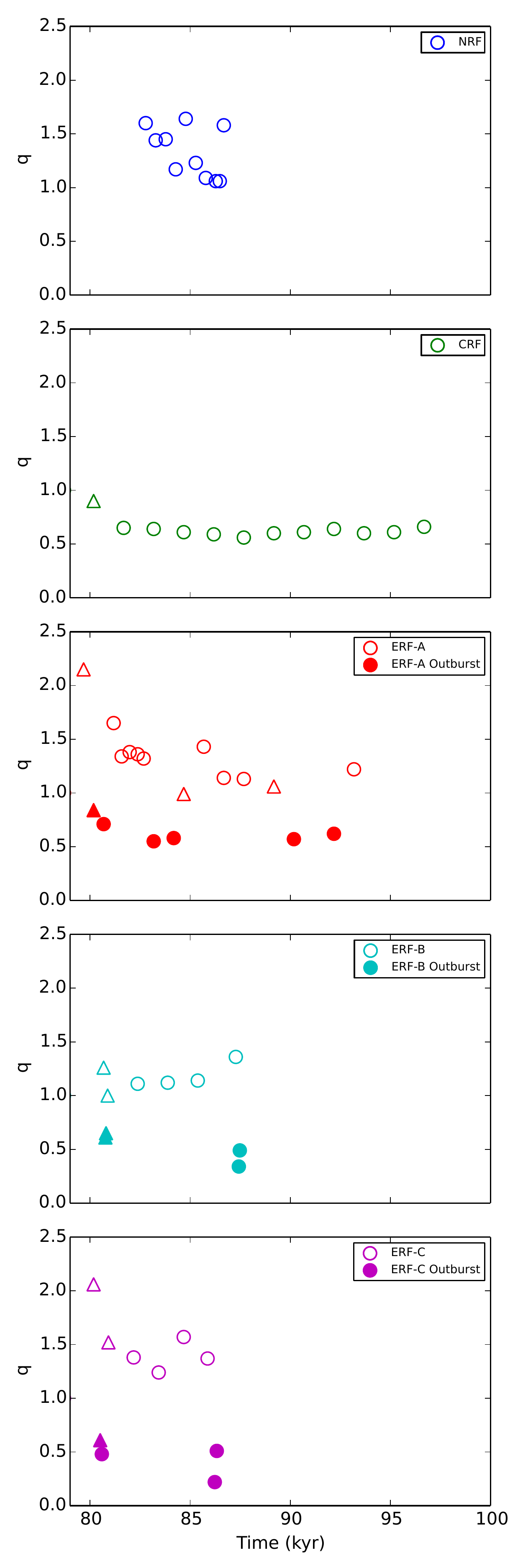}
 \caption{As per Fig.~\ref{fig:p_Time}, for the power exponent of the  temperature  profile  ($q$) evolution.}
 \label{fig:q_Time}
\end{figure}

 To quantify the differences in the disc structure for the different regimes of radiative feedback from the host protostar, we determine the radial dependence of the temperature, $T(r) \propto r^{-q}$,  and density, $\Sigma(r) \propto r^{-p}$, profiles as they are described by the exponents  $q$ and $p$. Fig.~\ref{fig:T_R} and Fig.~\ref{fig:sig_R} show representative fits to temperature and surface density profiles for the NRF, CRF and ERF-A snapshots (both during the quiescent and outburst phase). When fitting to each density and temperature profile, we set the radial range of the fit through visual inspection in order to ignore the inner flat region. There are  significant differences in the profiles between snapshots, so this process cannot be automated. This is the same procedure that is followed by observers when fitting profiles of observed discs.

We present the time evolution  of $p$ and $q$  in Fig.~\ref{fig:p_Time} and Fig.~\ref{fig:q_Time}. Since we are interested in the disc properties and not in disc fragmentation, we select snapshots before fragmentation happens (see Table~\ref{tab:ERFparams}). For the ERF models, snapshots that correspond to times in which episodic accretion is ongoing are shown as filled symbols. In determining $p$ and $q$, we use a sufficiently large radial extent for the fits.  However, this is difficult to do  for early times due to small disc extents, and when discs are undergoing GIs. $q$ and $p$ values that correspond to fits of radial extent $< 30 \ \text{AU}$ are denoted by triangles in Fig.~\ref{fig:p_Time} and Fig.~\ref{fig:q_Time}. Another uncertainty in the fits in some cases is due to the presence  of a central dense region where the density drops only slightly with radius (hereafter we call this region the disc ``bulge"). These uncertain values are also indicated by triangles.

The exponent of the surface density profile $p$ varies from 1.7 up to about 3 in the NRF and CRF models, whereas in the ERF models it  varies rather erratic from very low values in some in some cases ($\sim 0.8$) up to about 3 \citep[e.g. see][]{Kwon:2009a}. One would expect that due to inner disc gas being transported outwards during an outburst, the radial surface density profile would be more shallow, i.e. $p$ would decrease. After this gas redistribution, one would then expect $p$ to increase again. Although we note that there are indeed outburst events in which the characteristic $p$ reduction is observed, in general the evolution of $p$ after an outburst onset for all ERF models does not follow this pattern, probably due to the effect of spirals arms induced by GIs. Prior to an outburst, for all ERF models strong GIs are present, as it is evident from the presence of strong spirals. As the spirals are radially extended, the increased density in the outer disc produces a shallower radial density profile, with lower $p$ values compared to equivalent smooth morphology snapshots. Once an outburst occurs, the redistribution of material is expected to result in a decrease in the value of $p$. As the radially extended spiral structure already result in low $p$ values, the effect of material redistribution does not reduce $p$ greatly.
We should note here that one should be cautious when comparing  results from hydrodynamic simulations to observations of protostellar discs. The models presented here provide predictions for the gas distribution under different radiative feedback regimes from the young protostar. Observations in either infrared or (sub-)mm wavelengths trace  dust distribution in discs. The gas distribution may well be different from the dust distribution \citep[e.g.][]{tsukamoto16,Isella:2016a}.

\begin{table*}
	\caption{Average exponent values for radial surface density ($p$) and temperature ($q$) profiles. ERF snapshots are grouped according to their  morphology, i.e. the  presence  or lack of spiral structure (``GI" vs ``smooth", respectively). Examples of snapshots with GI and smooth morphology are presented in the left and right panels of Fig.~\ref{fig:AziProf_comp}, respectively. The average $q$ exponents  are evaluated in ERF models for both the quiescent phase (ERF) and the outbursting phase (ERF-Outburst).}
	\renewcommand{\arraystretch}{1.5} 
	\begin{tabular}{ c c c c c c c c c }
		\hline
		Morphology & \multicolumn{3}{c}{$p$} & & \multicolumn{4}{c}{q} \\
		\cline{2-4}
		\cline{6-9}
		 & NRF & CRF & ERF & & NRF & CRF & ERF & ERF-Outburst \\
		\hline
		GI & $2.3$ & $2.5$ & $2.2$ & & $1.3$ & 0.6 & $1.2$ & $0.4$ \\
		Smooth & - & $2.5$ & $2.2$ & & - & $0.6$ & $1.4$ & $0.6$\\
		\hline
	\end{tabular}
\label{tab:expcomp}
\end{table*}

The temperature profile exponent $q$ varies from 1~to~$\sim~1.6$ for the NRF model, is rather constant around 0.6 for the CRF model, and varies from 0.4 to 1.5 in the ERF models.  The variance  of $p$ and $q$ in the NRF model can be  attributed to the evolution of the disc as it accretes material asymmetrically from the infalling envelope and as GIs start to develop. The temperature profile is steeper than what it would be expected for a passive flared disc ($q=0.5$) but in this model  heating from the central star has not been taken into account. The NRF model can be directly compared the model of \citet{Tsukamoto:2015a}  that uses the flux-limited diffusion to treat the radiative transfer within the disc. They find that $p\sim1.65$ and $q=1.1$, which are consistent with our estimates (although they do not comment on the evolution of these values). In the CRF model the central star heats the disc so that $q$ does not drop below 0.5, similarly  to observations of T-Tauri discs \citep{andrews09}. Additional heating due to viscosity and/or transient spirals provide additional heating so that eventually $q\sim 0.6$. All three ERF models show significant variability, due to  the development of GIs in the quiescent phase, mass loading from the infalling envelope, and the role of highly variable radiative feedback  from the star  during and after the outburst phase. In the ERF runs, prior to the onset of an outburst, the temperature distribution is characterised by high values ($q>1 $), implying a sharp drop in temperature as a function of radius. During accretion outbursts, radiative feedback from the star heats the outer disc, resulting in a shallower temperature distribution ($q\sim0.5$) (see Fig.~\ref{fig:T_R}, bottom panel). After an outburst stops, the disc returns to the pre-outburst temperature profile until the next outburst event.

In Table~\ref{tab:expcomp} we present the average exponent values for all three radiative feedback models. We group the disc snapshots for each feedback regime depending on their morphology, defining a ``GI  morphology" when spiral structures are present in the surface density distribution, and a ``smooth morphology" for snapshots that show no significant spiral structure in the disc surface density.  We do not see  significant differences in the disc density and temperature distributions between different disc morphologies. We find that radial density profiles in the CRF case  have a slightly larger average $p$  than both the NRF and ERF equivalents, indicating that transient GIs are less efficient at redistributing angular momentum in the disc and changing its structure. The temperature gradient is much steeper in the NRF model ($q\sim1.3$) than in CRF model ($q\sim0.6$) demonstrating the  effect of protostellar radiative feedback in shaping the disc temperature profile. The temperature distribution in the ERF models is dual, resembling the NRF case during the quiescent phase and the CRF case during the outburst phase.

\section{Mass  estimates of the  protostar and its disc using position-velocity diagrams}\label{pv_analysis}

We showed in the previous section that the disc structure varies significantly during the early phases of its formation and evolution, depending on the type of radiative feedback from the central protostar. The disc dynamics at these early stages may therefore diverge from the dynamics of stable, axisymmetric discs, like e.g. the discs of T Tauri stars \cite{Hartmann:1998a}. Therefore, it is important to examine whether the protostellar masses derived from gas kinematics, i.e. the position-velocity (PV) diagram \citep{richer_padman91}, are reliable. Hydrodynamic simulations like the ones we present here provide the means to compare kinematic estimates of the mass with the actual mass of the  protostar-disc system.

\subsection{Defining the disc in simulations}

We adopt three criteria using azimuthally averaged  properties to define the disc radius around the protostar in the simulations. The first criterion defines the disc extent by evaluating the average azimuthal velocity $v_\phi(r)$ and comparing it to the Keplerian velocity, $v_\text{K}(r)$. We compute the Keplerian velocity using
\begin{equation*}
v_{K}(r) = \sqrt{\frac{G (M_\text{d,int}(r)+M_{*})}{r}} \ , 
\end{equation*}
 where $M_\text{d,int}(r)$ is the disc mass interior to $r$ and $M_{*}$ is the mass of the central protostar. The disc edge is defined as the radius where $v_\phi(r) \textless 0.9 v_{K}(r)$.  The second criterion defines the extent of the disc using the  radial infall velocity $v_{r}(r)$: we define the disc extent as the radius at which $v_{r}(r) \textgreater 2 \pi v_\phi (r)$. This  ensures the disc  material will orbit at least once around the central protostar before being accreted.  \citet{stamatellos12} find that both $v_{K}$ and $v_{r}$ disc criteria provide comparable disc masses and  radii, though the $v_{r}$ restriction produces systematically slightly smaller disc radii. The final criterion to determine disc extent and $M_\text{d}$ uses information of disc surface density, $\Sigma(r)$. We adopt a cut-off  value of $\Sigma(r) \textless \Sigma_\text{thresh}$ to define the disc radius. For our analyses, we adopt a fiducial value of $\Sigma_\text{thresh} = \ 20 \ \text{g cm}^{-2}$. The first two criteria identify rotationally supported discs, whereas the last criterion also identifies disc-like structures around the protostar.
 
 Once the disc radius is determined, a $z < 0.5 \ r$ restriction is imposed. This ensures that excess envelope emission is omitted when calculating $M_\text{d}$. The value of $M_\text{d}$ is then simply the mass interior to the computed disc radius, excluding the protostellar mass. The mass of the central protostar, $M_*$, is added to this mass to give a protostellar system mass, i.e. $M_\text{sys} = M_* + M_\text{d}$. 

\subsection{PV diagram construction from hydrodynamic simulations}

To mimic observational PV diagrams  we place the particles (that represent the gas in the SPH simulations) in projected spatial ($r$) and line-of-sight velocity  ($v_\text{\tiny{LOS}}$) bins. We use particles that are within 20~AU from the projected disc major axis, thus emulating a  slit with size $|\mathcal{H}| = 40 \ \text{AU}$. For each of the position-velocity bins we compute the flux at $\lambda = 1.3 \ \text{mm}$ (ALMA band 6), using the equation  \citep{hildebrand83}
\begin{align}
F_\lambda = B_\lambda(T_\text{d}) \kappa_\lambda \Sigma_\text{d} \Delta\Omega .
\end{align}
In the above equation, $ B_\lambda(T_\text{d})$ is the Planck function for a given dust temperature, $\kappa_\lambda$ is the opacity, $\Sigma_\text{d}$ is the dust surface density, and $\Delta\Omega$ is the solid angle of source. Assuming the gas and dust temperatures are equal, the dust temperature adopted for the Planck function is computed for every position-velocity bin, by taking the line-of-sight average temperature. This method is expected to more accurate, compared to the commonly adopted assumption of an isothermal disc (see \citealp{dunham14}). To calculate the flux we assume that the disc is optically thin. However this has two caveats: (a) in position-velocity bins where there is significant contribution from the envelope, the average temperature and hence the flux from the disc will be underestimated, and (b) for lines-of-sight intersecting the inner disc,  flux may be overestimated, as dense inner regions of the disc are expected even in (sub-) mm wavelengths to be optically thick. \citet{evans17} also notes that when computing flux, in the case of non-isothermal gravitationally unstable discs, the use of a constant $\kappa_\lambda$ may be a poor representation of the actual opacity, and therefore of the flux from the disc. Nonetheless, we adopt $\kappa_{\rm 1.3mm} = 2.3 \ \text{cm}^{2} \ \text{g}^{-1}$, as in \cite{Tobin:2016b}. $\Sigma_\text{d}$ is computed by using the gas surface density $\Sigma$ as computed from the simulation, and converting to $\Sigma_\text{d}$ assuming a constant dust-to-gas ratio of $0.01$. Finally, $\Delta\Omega = ( |\mathcal{H}| \Delta r ) / d^{2} $, where $d = 140 \ \text{pc}$, the assumed distance to the source, $\Delta r$ is the spatial bin-width, and $\mathcal{H}$ the assumed size of the slit.

Motivated by the ALMA spatial resolution in the Taurus star forming region \citep{alma15}, we set $ \Delta{r}=4 \ \text{AU}$  out to $|r| \le 100 \ \text{AU}$, and a velocity bin-width of  $\Delta v_\text{\tiny{LOS}} = 0.25 \ \text{km s}^{-1}$  up to   $|v_\text{\tiny{LOS}}| \le 10 \ \text{km s}^{-1} $. This ``resolution" is rather conservative when compared with the anticipated maximum resolution of ALMA\footnote{https://science.nrao.edu/facilites/alma/summary} ($\Delta v_\text{\tiny{LOS}} = 0.05 \ \text{km s}^{-1}$). We find that reducing this value (thus increasing the resolution) has no significant impact on the overall conclusions. 
 
 For each of the simulations presented here, the rotation of the disc  is anticlockwise, such that the PV diagrams are populated in upper right and lower left quadrants. The mass of the system, $M_\text{sys}$, is computed by fitting Keplerian profiles to the edge of position-velocity bins in the upper right and lower left quadrants. The edge of the respective quadrant is defined when a bin with non-zero surface density is found. We fit to the edge, as opposed to the location of maximum emission (as per \citealp{jorgensen09,tobin12,ohashi14}). This is motivated by the results of \citet{seifried:2016a} in which fits to the PV quadrant edge are found to more faithfully recover $M_\text{sys}$, whereas fitting by the maximum tends to underestimate $M_\text{sys}$. The maximum radius to which we fit the Keplerian profile is set by the radius in either quadrant at which $\Sigma < 20 \ \text{g cm}^{-2}$. For each quadrant we then compute $M_{\text{sys}}$ by fitting a Keplerian profile $v_{K} (r) = \sqrt{GM_\text{sys} / r}$ to the data. The final estimate for $M_{\text{sys}}$ is the average of the masses computed from both quadrants of the PV diagram.

\begin{figure*}
 \includegraphics[width=150mm,keepaspectratio]{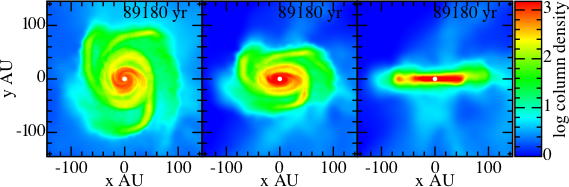}
 \caption{Surface density (in ${\rm g\ cm}^{-2}$) of an episodic radiative feedback model (ERF-A) snapshot at $89.1 \ \text{kyr}$, for different inclination angles: $0^{\circ}$ (left panel), $i = 60^{\circ}$ (middle panel) and $i = 90^{\circ}$ (right panel). }
 \label{fig:spatial_inc}
\end{figure*}

\begin{figure*}
 \includegraphics[width=0.95\columnwidth]{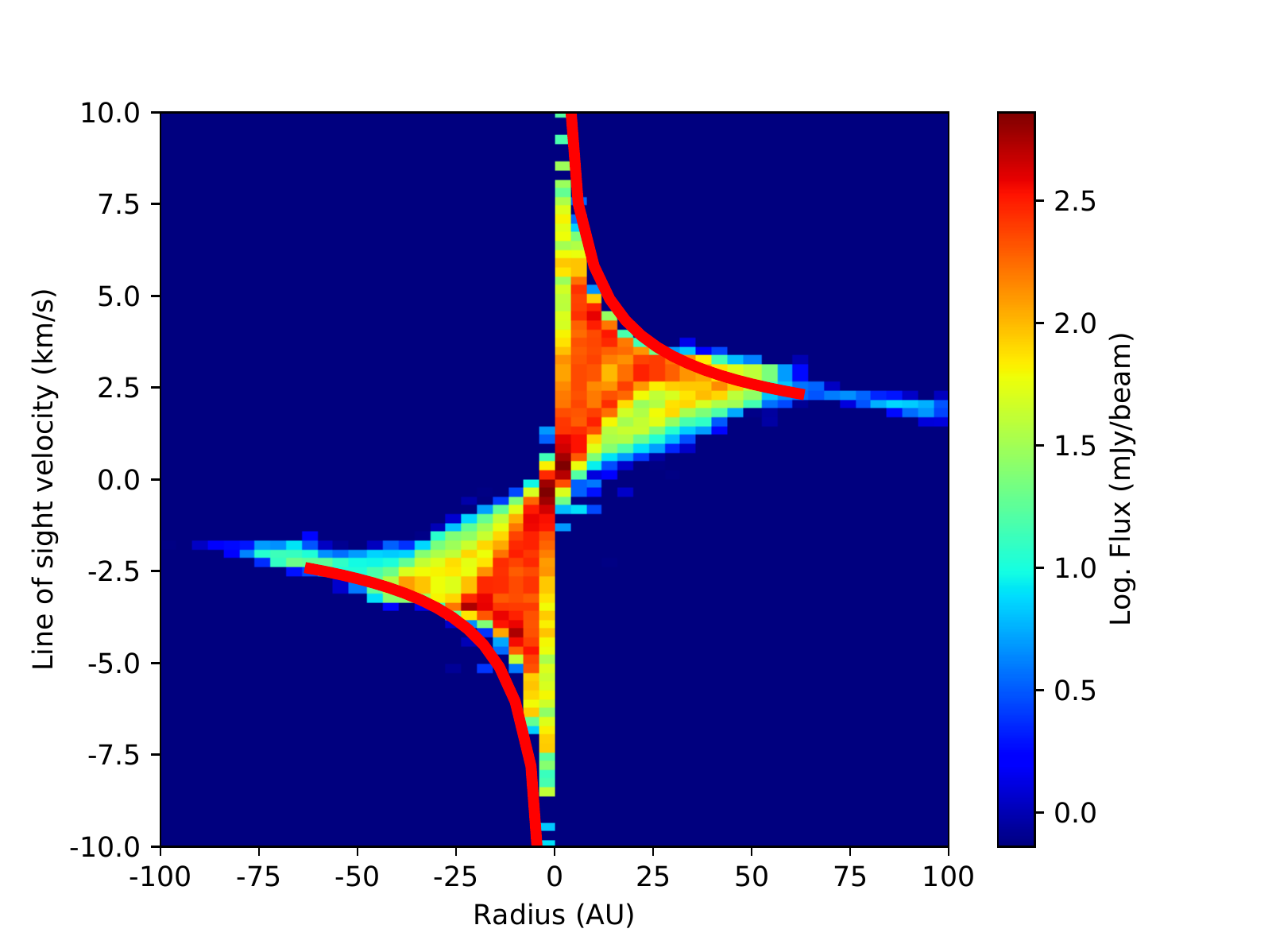}
 \includegraphics[width=0.95\columnwidth]{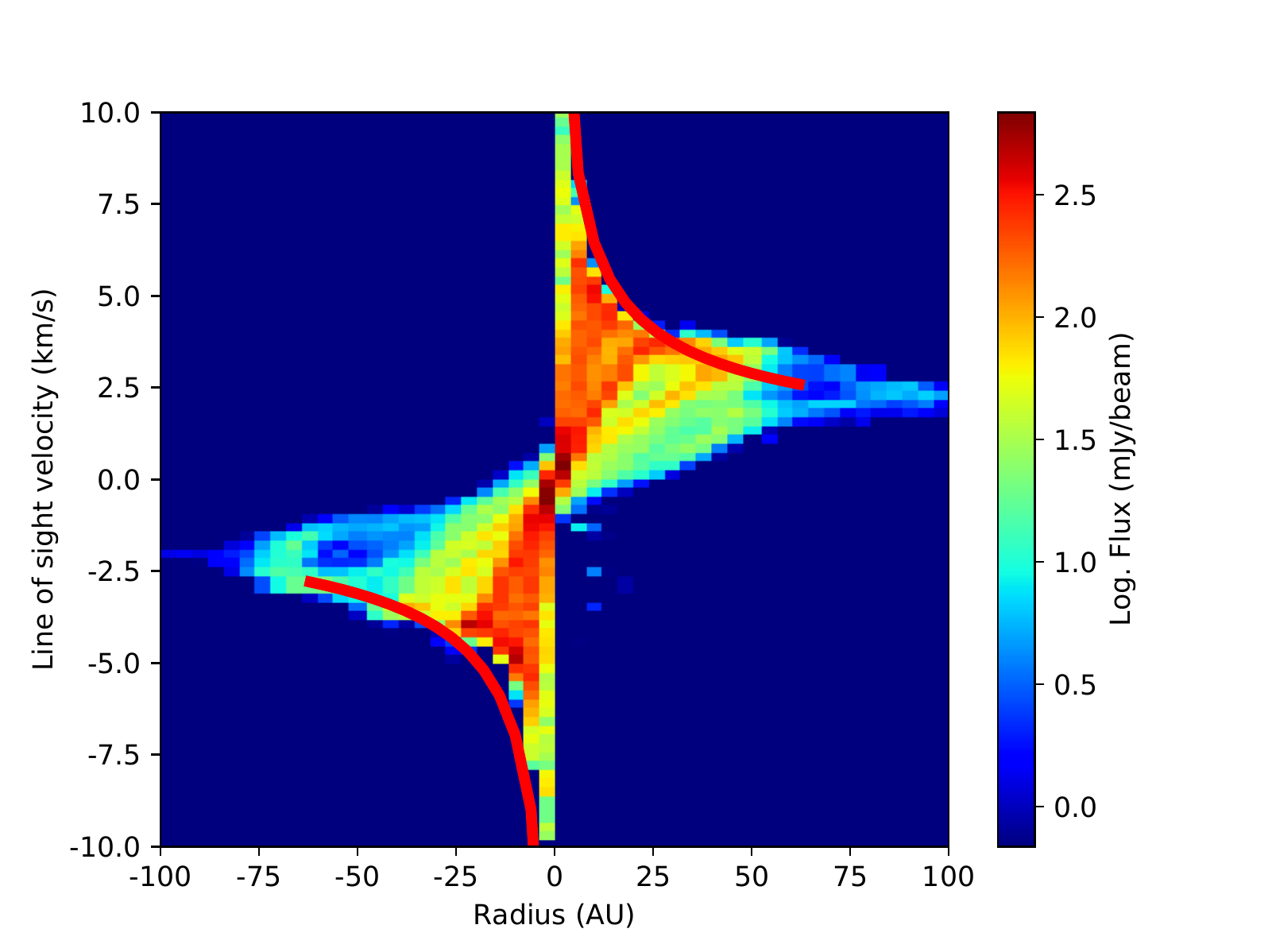}
 \caption{PV diagrams for the snapshot  of Fig.~\ref{fig:spatial_inc} (left panel: $i = 60^{\circ}$; right panel $i = 90^{\circ}$). The fitted Keplerian velocity profile is overplotted as a red curve.}
\label{fig:PV_incl}
\end{figure*}

The PV diagram of a specific system depends on the orientation of the disc with respect to the observer. For the following analysis, we adopt two inclinations, $60^{\circ} \text{ and } 90^{\circ}$ (phase-on discs), motivated by recent results suggesting that masses estimated through Keplerian fits to PV diagrams are significantly inaccurate for $i < 60^\circ$ \citep{seifried:2016a}. Surface density plots for both inclination cases, and also when $i = 0^\circ$ as reference, are illustrated in Fig.~\ref{fig:spatial_inc} with corresponding PV diagrams plotted in Fig.~\ref{fig:PV_incl}. The PV diagrams of Fig.~\ref{fig:PV_incl} have the Keplerian fits to the kinematic data overplotted as red solid curves. In both Fig.~\ref{fig:spatial_inc} and Fig.~\ref{fig:PV_incl} the outburst-phase snapshot of the ERF-A model (see Fig.~\ref{fig:spatial}) is used. 

\begin{figure*}
 \centering
   \subfigure[NRF]{\includegraphics[width=85mm,keepaspectratio]{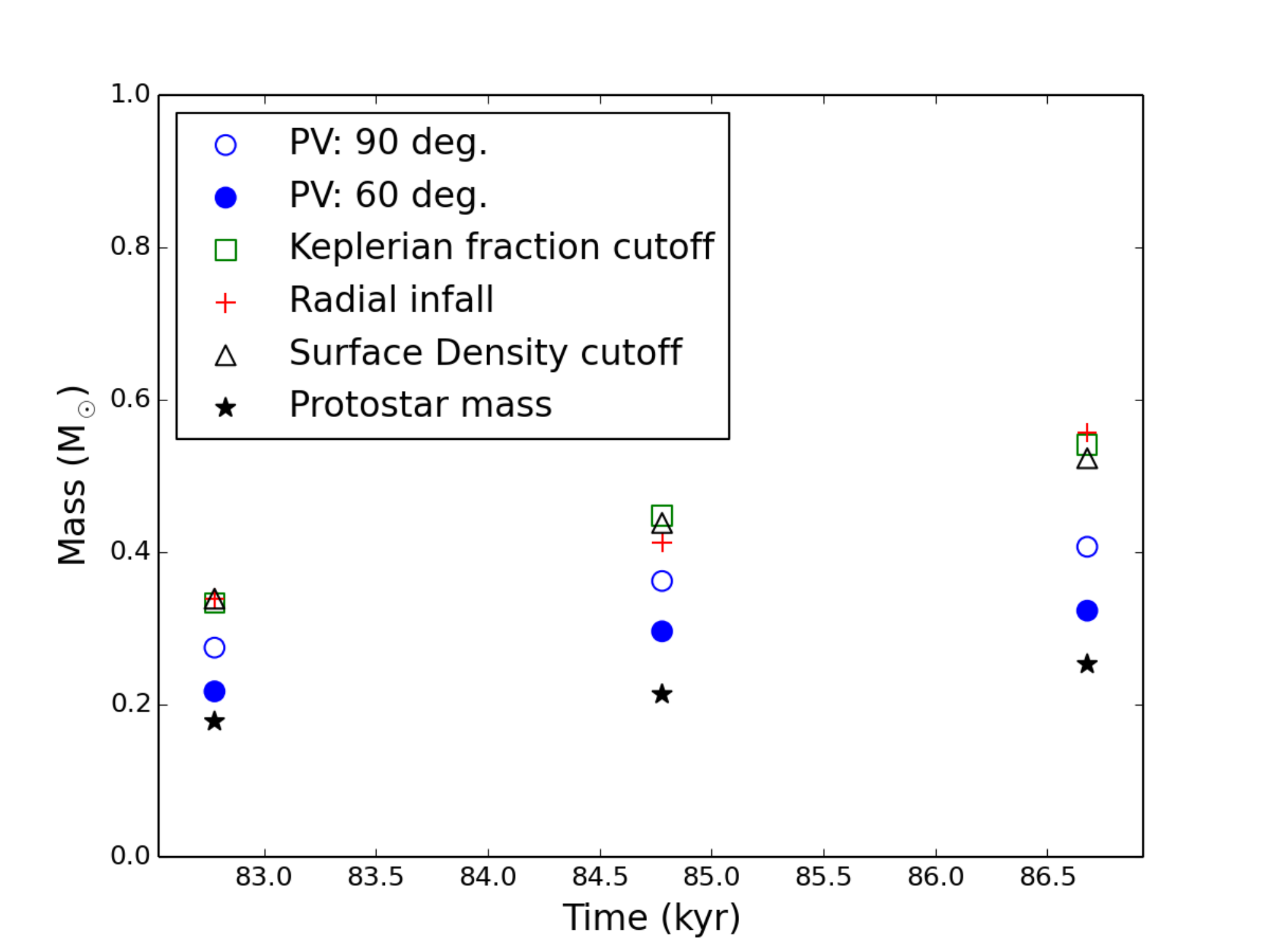}}
   \subfigure[CRF]{\includegraphics[width=85mm,keepaspectratio]{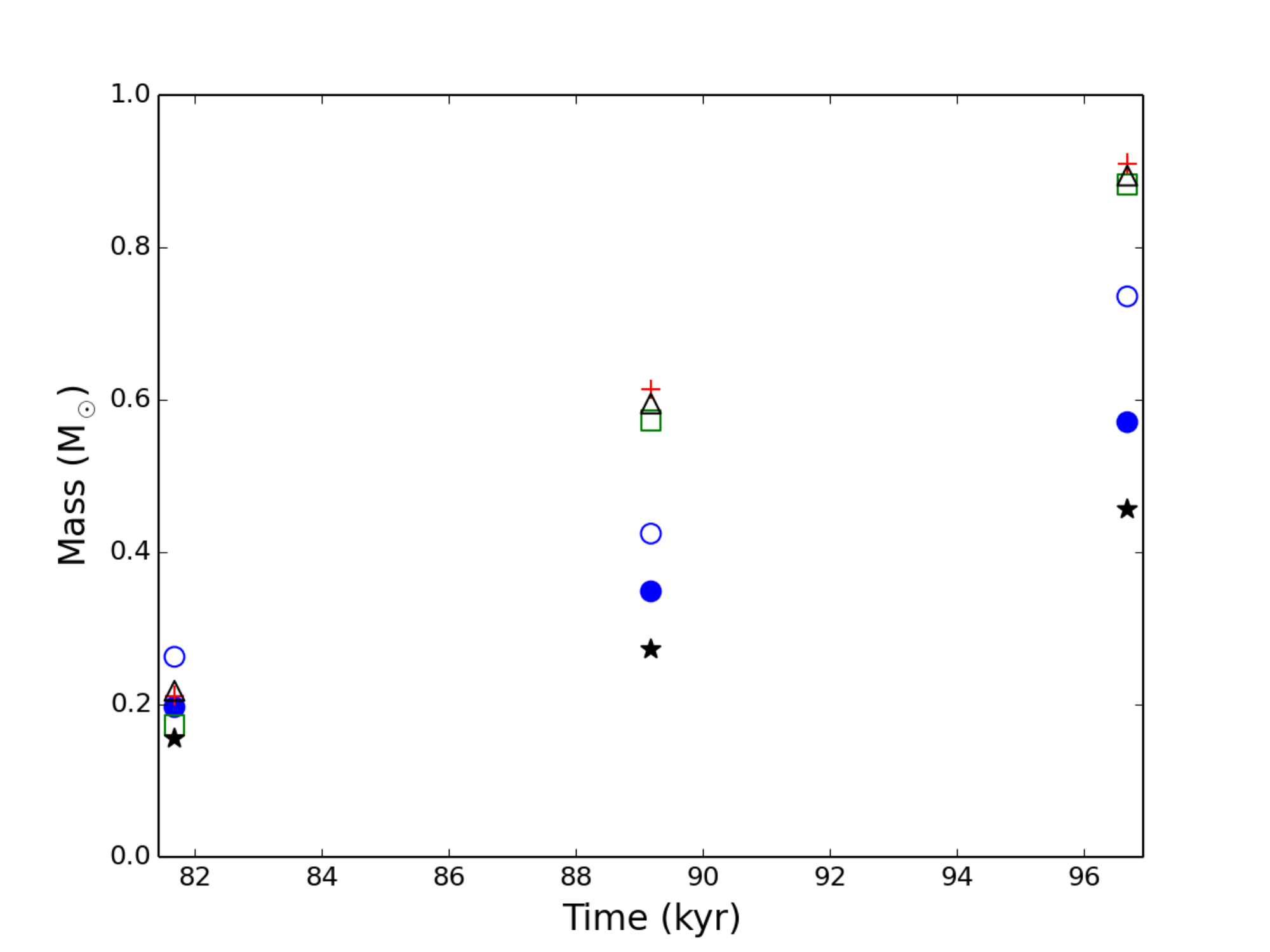}}
   \subfigure[ERF-A]{\includegraphics[width=85mm,keepaspectratio]{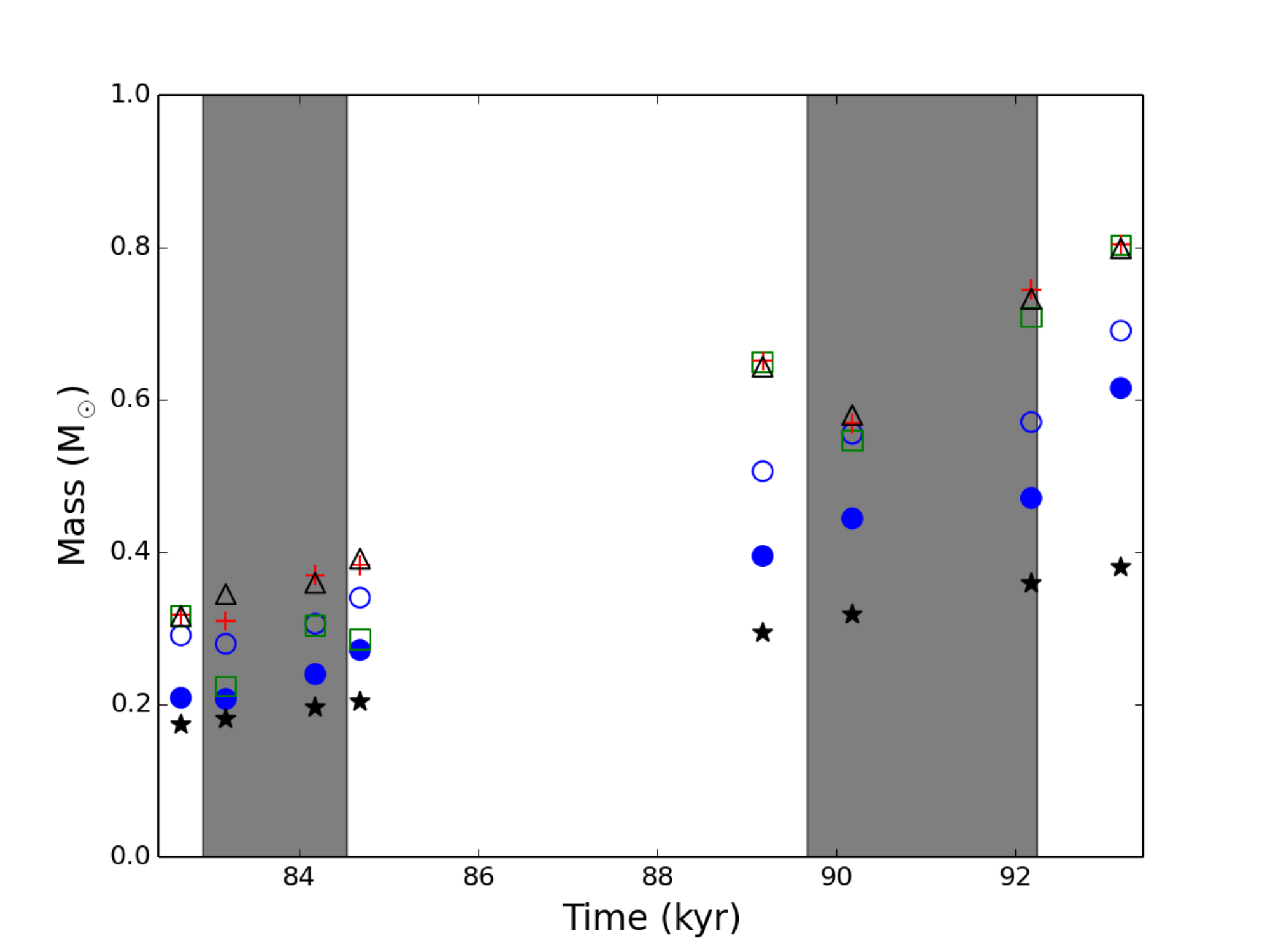}}
   \subfigure[ERF-B]{\includegraphics[width=85mm,keepaspectratio]{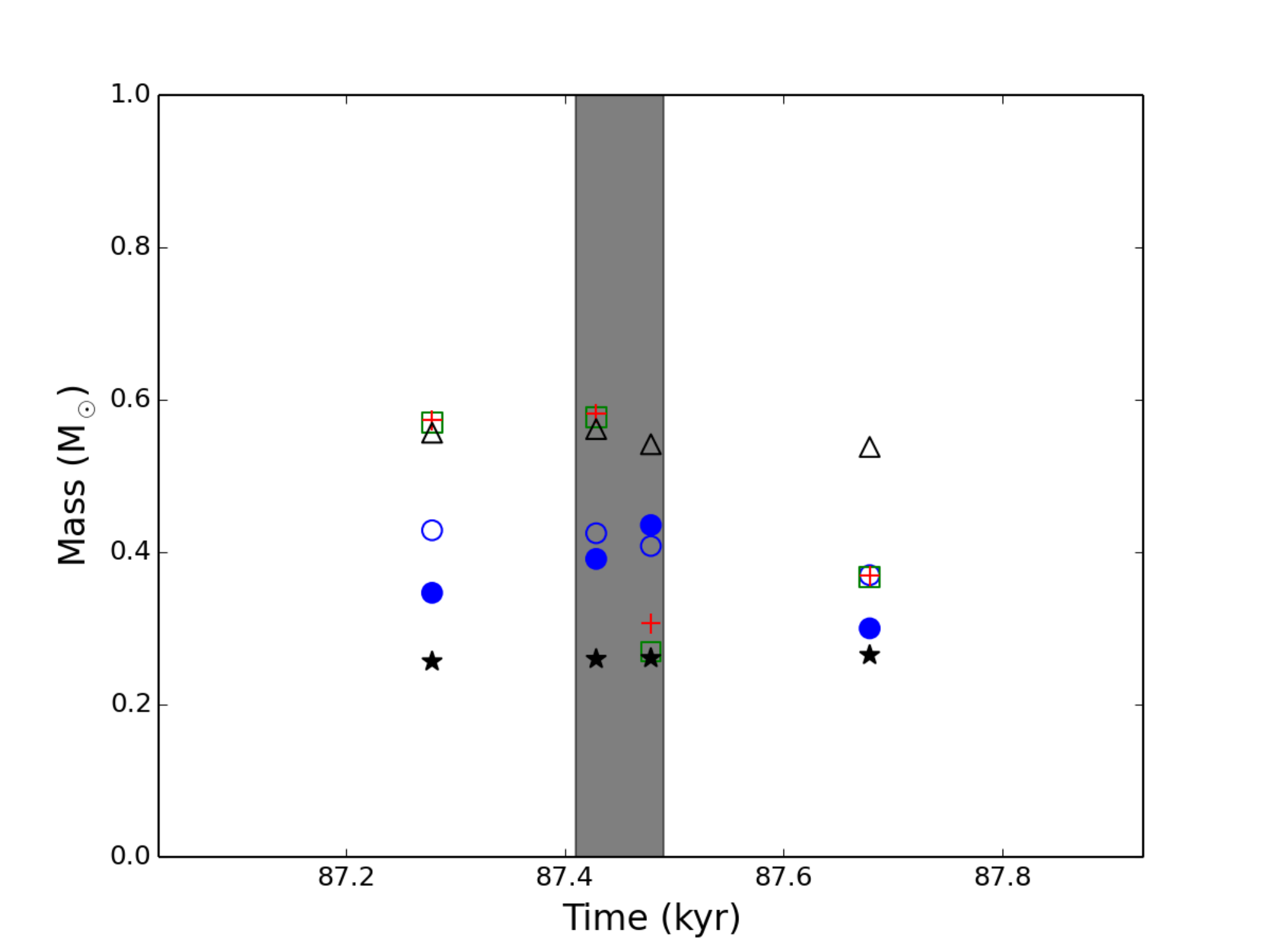}}
   \subfigure[ERF-C]{\includegraphics[width=85mm,keepaspectratio]{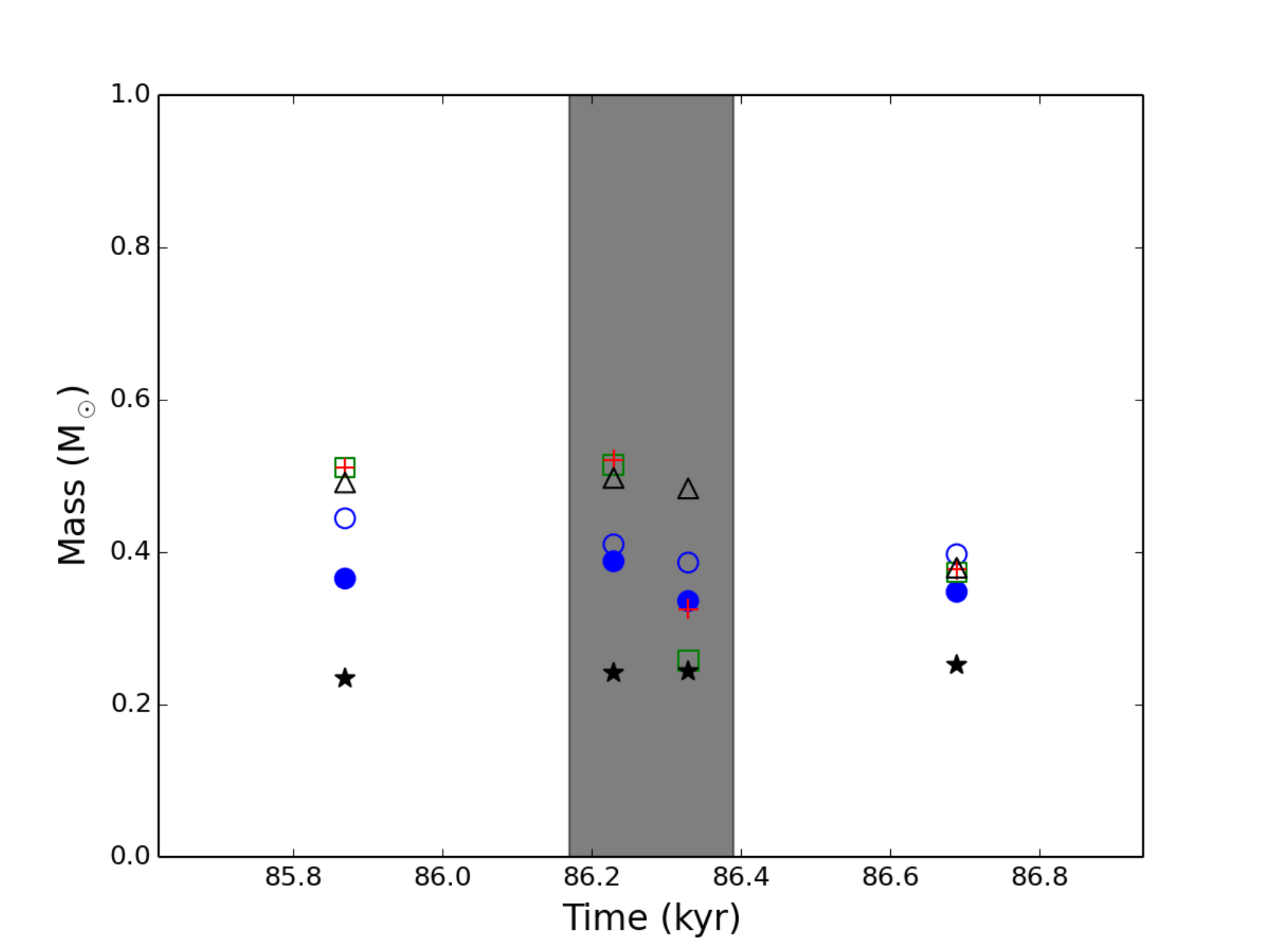}}   
 \caption{The evolution of protostellar and disc mass ($M_\text{sys}$) for the no radiative feedback (NRF), the continuous radiative feedback (CRF) and the three episodic radiative feedback (ERF) runs. Each panel includes mass estimates through position-velocity diagrams, for $i=90^\circ$ (open blue circles) and $i=60^\circ$ (filled blue circles). The actual $M_\text{sys}$ from the simulations are also plotted, as calculated from three different criteria to define the disc: the deviation from the Keplerian velocity $v_K$ (green open squares), the radius restricted by the radial velocity $v_r$ (red plus symbols), and the surface density threshold (open black triangles) criteria (see text). Black stars represent the protostellar mass, $M_*$ as  determined from the simulations. In the ERF panels the shaded regions represent the period where episodic accretion is ongoing.}
 \label{fig:mass_Time}
\end{figure*}

\subsection{How well do position-velocity diagrams estimate the mass of a protostellar system?}

We now compare the mass derived from kinematic considerations using PV diagrams  and the actual system mass derived from the simulations, $M_\text{sys}$, to test the accuracy of PV diagrams in estimating disc masses, especially in the case when an outburst happens, where the disc dynamics may deviate from a Keplerian disc. In Fig.~\ref{fig:mass_Time}, we present a series of  $M_\text{sys}$ evolution plots, for all runs. Each panel includes information on the PV diagram mass estimates ($i = 90^\circ$: open blue circles; $i = 60^\circ$: filled blue circles) and the masses derived from the hydrodynamic simulations. Each panel in Fig.~\ref{fig:mass_Time} also presents the protostellar mass (black stars). The difference between this mass and  $M_\text{sys}$ value for each snapshot  represents the disc mass. For each ERF run, we are particularly interested in the impact of time-varied radiative feedback on estimates on both the mass derived from the simulations  and kinematic mass derived from the PV diagram. As such, we include a snapshot before onset, after onset, before offset, and after offset for an accretion event.  For the ERF runs, the time interval at which the protostellar system is undergoing an outburst event is indicated by the shaded region. 
 
In the no radiative feedback and continuous radiative feedback runs (Fig.~\ref{fig:mass_Time}a,b), the values of $M_\text{sys}$ that are computed from the simulations using different criteria for defining the disc agree well for all snapshots as there is no radiative feedback variability. However,  we note that the $M_\text{sys}$ estimates from the simulations are systematically higher than the PV inferred counterparts. We also see that the mass estimates using the PV diagram are lower for increasing inclination. This is because the peak line-of-sight velocity scales with $\sin (i)$ and, since $M_\text{sys} \propto v_\text{K}^{2}$, the mass estimate scales with $\sin^{2} (i)$. Using this relation, we expect that $M_\text{sys}(60^\circ) \sim 0.75 \ M_\text{sys}(90^\circ)$. To a few percent, we find that the PV mass estimates in NRF and CRF models are consistent with this analysis. Therefore, it is needed to take into account the disc orientation when estimating the mass of the protostellar system (as noted by e.g. \citealp{hara:2013a, murillo:2013a}).

\begin{figure*}
 \centering
   \subfigure[Before onset]{\includegraphics[width=82mm,keepaspectratio]{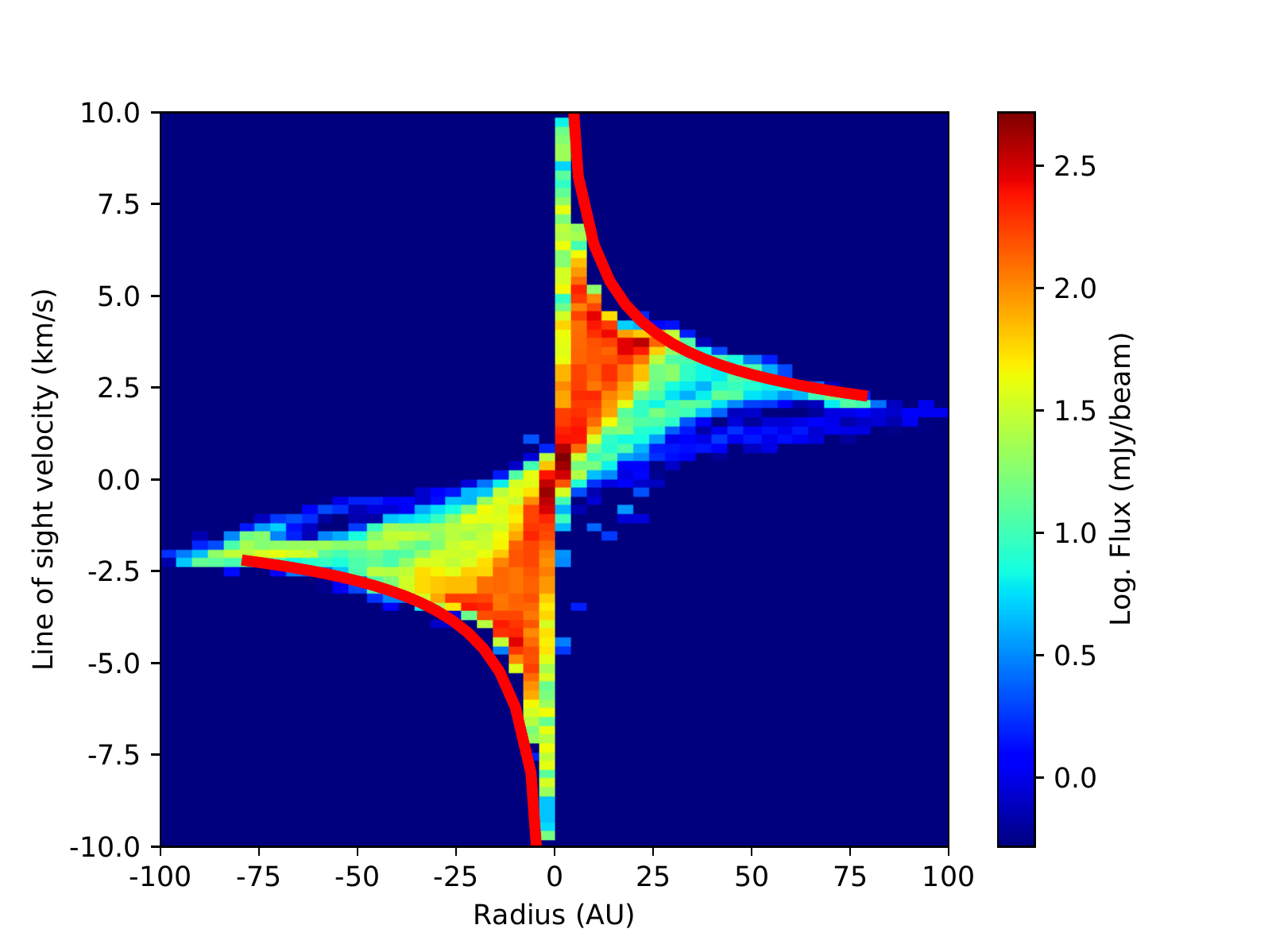}}
   \subfigure[After onset]{\includegraphics[width=82mm,keepaspectratio]{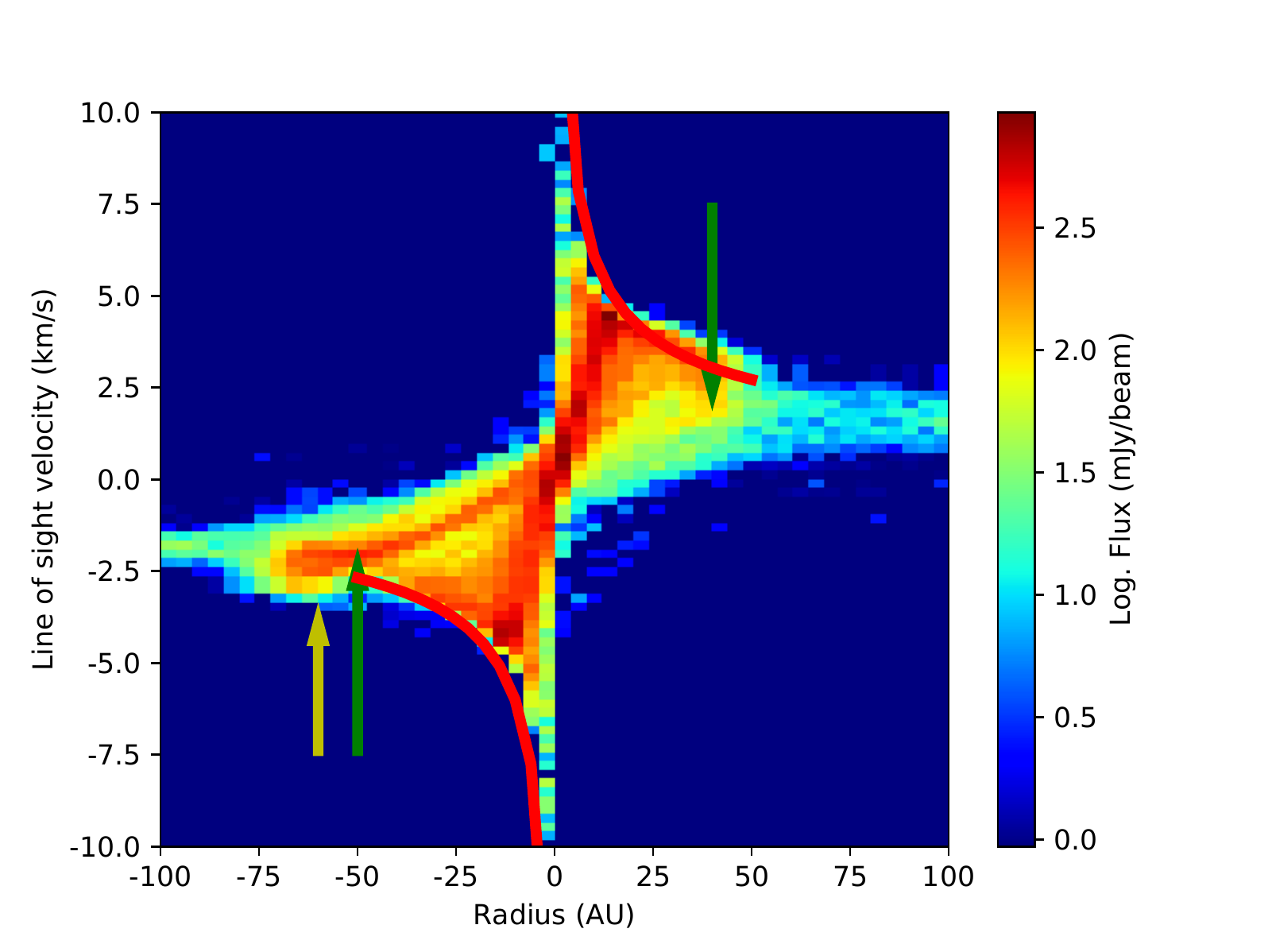}}
   \subfigure[Before offset]{\includegraphics[width=82mm,keepaspectratio]{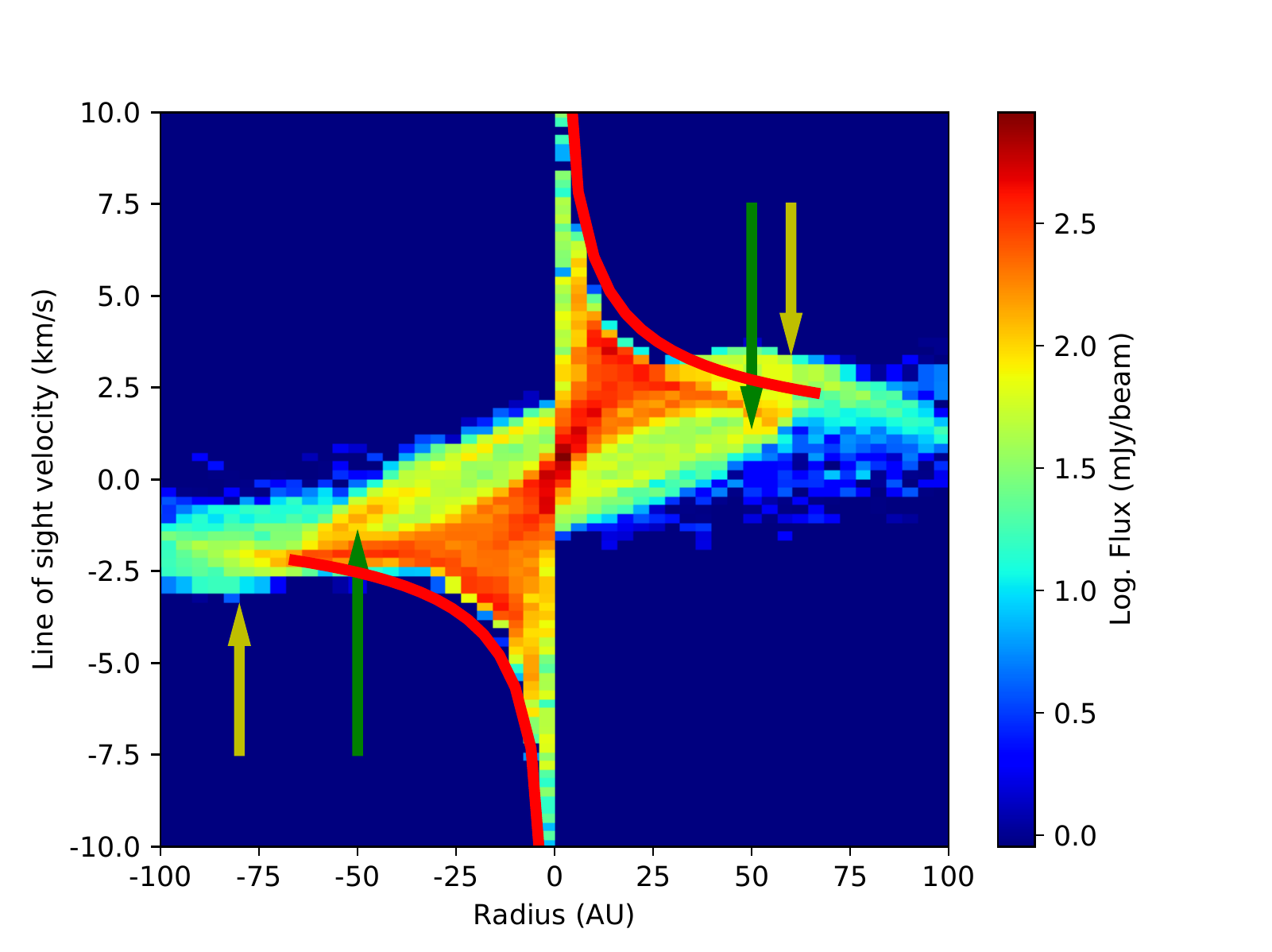}}
   \subfigure[After offset]{\includegraphics[width=82mm,keepaspectratio]{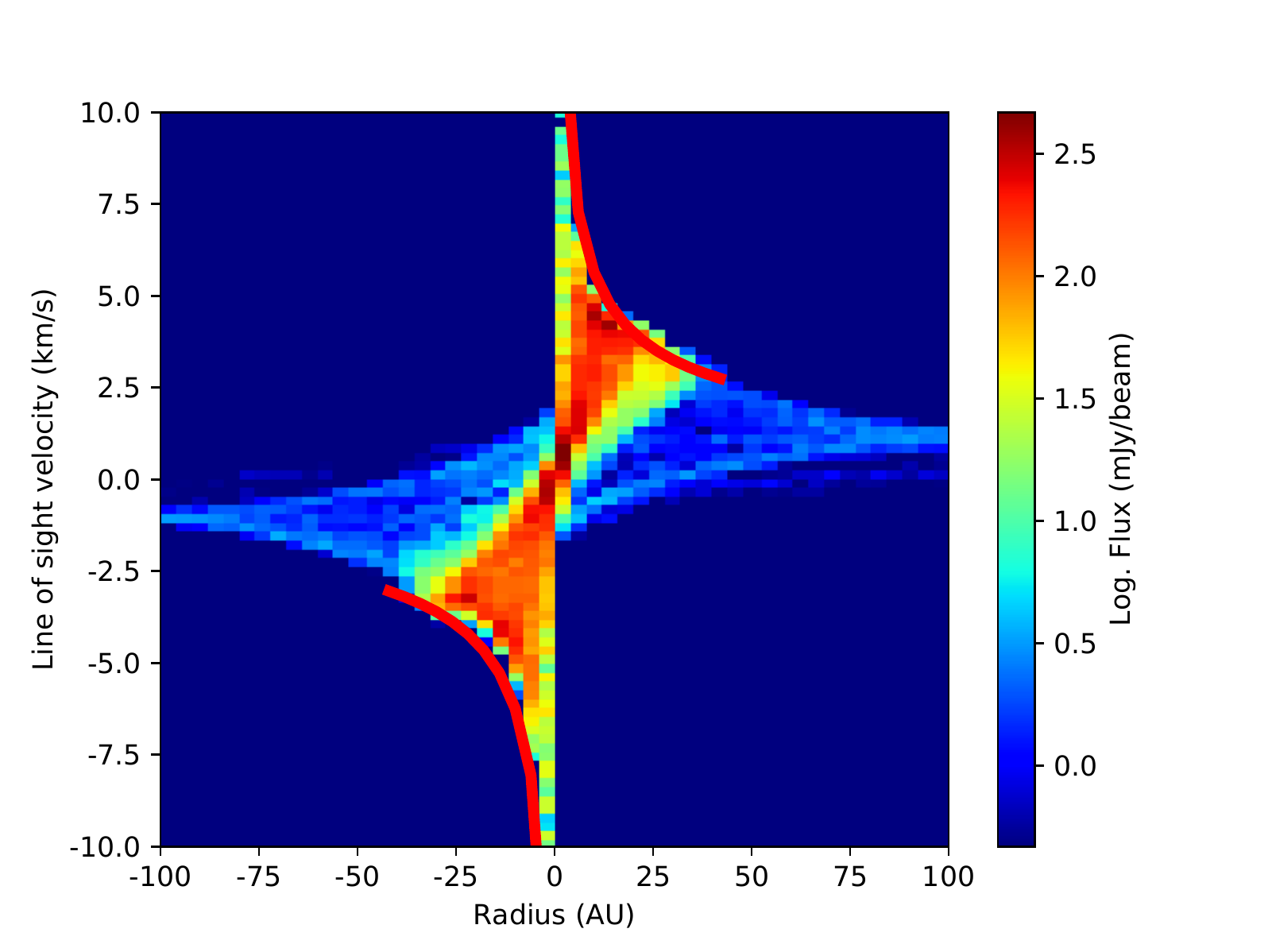}}
 \caption{PV diagrams for the episodic radiative feedback run C (ERF-C). Panels (a)-(d) correspond to snapshots before the onset, after the onset, before the offset and after the offset of an episodic accretion event. The PV diagram changes due to  the impact of episodic radiative feedback on the kinematics of the protostellar disc. Locations at which spirals are shown as fingers in each PV diagram are indicated by green arrows. Super-Keplerian velocities due to radiative feedback in (b) and (c) are indicated by yellow arrows. Red lines represent fitted Keplerian profiles, to highlight super-Keplerian material during/after outburst.}
 \label{fig:PV_ERF}
\end{figure*}

The PV diagrams for the ERF runs demonstrate significant variability.  In Fig.~\ref{fig:PV_ERF} we provide a series of PV diagrams for an outburst event in the ERF-C run at $86.2 \ \text{kyr}$ ($i=90^\circ$). Once an episodic accretion event begins, the PV diagram evolves due to the radiative feedback impacting on both the kinematic and thermal properties of the disc.  We find that the impact of episodic radiative feedback may result in disagreement between simulation $M_\text{sys}$ estimates using different criteria. This is because the dynamics of the system are strongly affected by the outburst so that there is no rotationally-supported disc but rather a disc-like structure. In the case of the ERF-A run (Fig.~\ref{fig:mass_Time}c), in which  outbursts events are longer but mild,  we find that the masses estimated from the simulation using different criteria disagree but not as much as in  ERF-B and ERF-C runs, where the outburst events are  more intense with shorter duration. In the latter two cases, the effect of the radiative feedback from the outburst strongly influences the dynamics of the system. 

We find that the masses inferred through Keplerian fits in PV diagrams are systematically lower than  the actual masses of  protostellar systems. For $i = 90^\circ$ orientations, the $M_\text{sys}$ value from the PV Keplerian fits are  $\sim20$\% less than the actual system mass.  The mass of the protostar itself cannot be directly inferred from PV diagrams for early-phase discs.  At this stage the disc mass is a significant fraction of the mass of the protostar and therefore the contribution of the disc self-gravity to the Keplerian velocity of the gas  cannot be ignored.

\subsection{Features in PV diagrams indicative of spirals}

 \citet{douglas13} have found that PV diagrams may be used to indicate the presence of spiral structures due to  gravitational instabilities. They have pointed out to  ``fingers'' in the PV diagrams, i.e. significant, radially extended over densities. In this work we confirm this finding (see Fig.~\ref{fig:PV_ERF};  in each panel, the  fingers due to  spirals are denoted by the green arrows). The prominence of the ``fingers'' in Fig.~\ref{fig:PV_ERF} is dependent on the stage on accretion. GIs are expected to be more prominent prior to outburst (panel (a)), however due to inadequate outer disc flux contribution, the contrast is insufficient for identification. In panels (b)-(c), with increased disc heating, the kinematic features become more prominent. These ``fingers'' are however transient in nature,  with increased spiral flux contribution after outburst being disrupted as $Q$ increases, stabilising the disc. In panel (d) of Fig.~\ref{fig:PV_ERF}, no strong over densities are present, i.e. the radiative feedback has disrupted the spiral structure, leading to an axisymmetric disc surface density profile.

\subsection{Signatures of outbursts in  PV diagrams}

The series of PV diagrams in Fig.~\ref{fig:PV_ERF} tenuously indicates the accretion state of the protostellar system, i.e. whether a system is quiescent or undergoing an outburst). In panel (a) prior to the outburst onset, the outer disc is cool and spirals are present. The edge of each PV diagram quadrant may be represented by a Keplerian velocity profile. In panels (b) and (c) however, the impact of the increased radiative feedback results in significant disruption of disc kinematics, resulting in substantially super-Keplerian rotation at certain radii. This super-Keplerian gas, denoted in Fig.~\ref{fig:PV_ERF} by yellow arrows, is representative of the impact of outward flows of inner disc material due to strong radiative feedback. Once the episodic accretion event has stopped, the outward radial flows stop. The disc then returns to the Keplerian rotation profile, but without  any spiral structure due to the impact of the increased heating due to the preceding outburst.

\subsection{Radial pressure gradients in deeply embedded protostellar discs}

There are systematic differences between the mass derived from kinematics, as they are depicted in the PV diagram,  and the actual mass of the protostellar system.  We refer to the impact of the radial pressure gradient on modifying the gas orbital velocity from the Keplerian one, so that, at radius $r$, the orbital velocity of the gas, $v_\phi$, is 
\begin{equation}
v_\phi = v_K (1 - \eta)^{1/2}
\end{equation}
where $ v_K$ is the Keplerian velocity, $\eta = (p + q + 1) (h/r)^{2}$, and $h(r)/r \equiv H$ is the disc scale-height \citep[see][]{Armitage:2007a}. For a typical T Tauri disc in which $p = 3/2$, $q = 3/4$ and $H = 0.05$, $v_\phi \sim 0.997$. For early-phase discs, as the ones in the models presented in this paper, we have $H \sim 0.2$ (considerably thicker discs, e.g see \cite{Sakai:2017a}, due to high disc-to-stellar mass ratio and/or due to high protostellar accretion luminosities) and $(p+q+1)\sim4$, therefore $v_\phi \sim 0.917 \ v_K$, i.e 8\% less than the Keplerian velocity. Recalling that the  system mass  as computed from the PV diagram is $M_\text{sys} \propto v_K^{2}$, we find that Keplerian fits in PV diagrams underestimate the mass of a young disc by  $\sim 15\%$, which is similar to what we find in the simulations we have analysed here ($\sim 20\%$).  If embedded discs are even thicker  (say $H \sim 0.3$) then the disc mass  is underestimated even more (by $\sim35\%$). Therefore, previous studies of young embedded discs based on PV diagrams may  significantly underestimate disc masses.

 The role of increased radial pressure gradient in the case of early-phase discs is expected to play a major role in governing the dust dynamics \citep{weidenschilling77}. As dust grains that are decoupled from the gas (i.e. relatively large grains) do not experience a radial pressure gradient, their orbital velocity can be approximated as Keplerian. With an increased difference between dust and gas orbital velocity, the dust will therefore feel a stronger headwind from the gas. This headwind results in a loss of angular momentum and inward migration of dust towards the central protostar. Our results suggest that as the radial pressure gradient is stronger in young, embedded protostellar discs in comparison to later stage discs, the rate of dust depletion may be significant early-on during the disc evolution. As a result, the dust-to-gas mass ratio in young planet forming discs may be smaller than the one in the interstellar medium ($\sim 0.01$), which may  also lead to underestimating  disc masses by  $20-30\%$ when dust continuum observations are used \citep{birnstiel09,birnstiel10,tsukamoto16}.
 
 \section{Conclusions}\label{conclusions}

We have investigated the properties of young discs as they are still forming in collapsing molecular clouds, using hydrodynamic simulations that employ different radiative feedback from the protostar hosting the disc: no radiative feedback, with accretion-powered continuous radiative feedback  and with  episodic feedback. In addition, we have examined whether PV diagrams, commonly used by observational studies, can accurately determine the mass of  a young protostar and its disc. The key results are:

\begin{enumerate}[(a)]

\item The radial surface density and temperature profiles vary significantly as mass falls on the disc from the envelope. The profiles are particularly sensitive to the radiative feedback from the protostar, especially in the case when this is episodic. 

\item
The exponent of the surface density profile $p$ varies from 1.7 up to about 3 in the non-radiative feedback and in the continuous radiative feedback models. In the episodic radiative feedback models it  varies  erratically  from  $\sim 0.8$ up to about 3. The temperature profile exponent $q$ varies from 1 to $\sim 1.6$ in the non radiative feedback model, is rather constant around 0.6 for the continuous radiative transfer  model, and varies form 0.5 to 1.5 in the episodic radiative feedback models. We find that  the disc temperature drops faster with radius when the  radiative feedback from the protostar is weak. Shallow temperature profiles  ($q \sim 0.5$) are associated with outbursting embedded protostars.

\item PV diagrams may be used to provide important information about the structure of discs whose midplane is at high inclinations with respect to the observer (especially nearly edge-on discs).  Spiral arms  indicative of gravitational instabilities can not be seen in images of such discs but  can be inferred from the presence of characteristic structures (fingers) in the PV diagram. 

\item Super-keplerian velocities in the PV diagram may be indicative of a protostellar system that is undergoing an outbursting event, similar to FU Ori events at late-phase, T Tauri discs.

\item The mass of the protostellar system (i.e. the protostar and its disc) is underestimated  by  20\% when PV diagrams are used.  This is because young embedded discs are relatively thick and the radial pressure gradient reduces the orbital velocity below the Keplerian velocity of the gas by a factor of 8\%. Moreover, for the early-phase discs that we study here, the mass of the disc is significant and therefore should be taken into account when calculating the mass of the protostar from the PV diagram.
 
\item The significant pressure gradient expected in early-phase discs suggests that the rate of dust depletion due to  gas drag may be high in the deeply embedded phase. Therefore,  the  dust-to-gas mass ratio of young embedded discs may be lower than that of the ISM.

\end{enumerate}

\section*{Acknowledgements}

We thank the anonymous referee for insightful comments that helped improving the manuscript. We would also like to thank Stewart Eyres and Chang Won Lee for useful suggestions. BM is supported by STFC grant ST/N504014/1. DS is partly supported by STFC grant ST/M000877/1. Fig.~\ref{fig:spatial} and Fig.~\ref{fig:spatial_inc} have been generated using the visualisation software  SPLASH \citep{price07}.



\bibliographystyle{mnras}

\bibliography{macfarlane17} 


\bsp	
\label{lastpage}
\end{document}